\documentclass[fleqn,usenatbib]{mnras}

\usepackage{newtxtext,newtxmath}

\usepackage[T1]{fontenc}

\DeclareRobustCommand{\VAN}[3]{#2}
\let\VANthebibliography\thebibliography
\def\thebibliography{\DeclareRobustCommand{\VAN}[3]{##3}\VANthebibliography}

\usepackage{graphicx}	
\usepackage{amsmath}	

\usepackage{appendix}

\usepackage{enumitem}

\title[Clump properties with JWST/NIRCam]{Near-IR clumps and their properties in high-z galaxies with JWST/NIRCam}

\author[B. S. Kalita et al.]{
Boris S. Kalita,$^{1,2,3}$\thanks{E-mail: boris.kalita@ipmu.jp; kalita.boris.sindhu@gmail.com}\thanks{Joint-Kavli Astrophysics Fellow}
John D. Silverman,$^{1,4,5,3}$
Emanuele Daddi,$^{6}$
Wilfried Mercier,$^{7}$
\newauthor 
\,\,Luis C. Ho,$^{2,8}$
and Xuheng Ding$^{1}$
\\ \\
$^{1}$Kavli IPMU (WPI), UTIAS, The University of Tokyo, Kashiwa, Chiba 277-8583, Japan \\
$^{2}$Kavli Institute for Astronomy and Astrophysics, Peking University, Beijing 100871, People{\textquotesingle}s Republic of China\\
$^{3}$Centre for Data-Driven Discovery, Kavli IPMU (WPI), UTIAS, The University of Tokyo, Kashiwa, Chiba 277-8583, Japan\\
$^{4}$Department of Astronomy, School of Science, The University of Tokyo, 7-3-1 Hongo, Bunkyo, Tokyo 113-0033, Japan\\
$^{5}$centre for Astrophysical Sciences, Department of Physics \& Astronomy, Johns Hopkins University, Baltimore, MD 21218, USA\\
$^{6}$Universit{\'e} Paris-Saclay, Universit{\'e} Paris Cit{\'e}, CEA, CNRS, AIM, Paris, 91191,France\\
$^{7}$Aix Marseille Univ, CNRS, CNES, LAM, Marseille, France\\
$^{8}$Department of Astronomy, School of Physics, Peking University, Beijing 100871, People{\textquotesingle}s Republic of China
}

\date{Accepted 2025 January 7. Received 2025 January 3; in original form 2024 February 2}

\pubyear{----}

\begin{document}
\label{firstpage}
\pagerange{\pageref{firstpage}--\pageref{lastpage}}
\maketitle

\begin{abstract}
Resolved stellar morphology of $z>1$ galaxies was inaccessible before JWST. This limitation, due to the impact of dust on rest-frame UV light, had withheld major observational conclusions required to understand the importance of clumps in galaxy evolution. Essentially independent of this issue, we use the rest-frame near-IR for a stellar-mass dependent clump detection method and determine reliable estimations of selection effects. We exploit publicly available JWST/NIRCam and HST/ACS imaging data from CEERS, to create a stellar-mass based picture of clumps in a mass-complete sample of 418 galaxies within a wide wavelength coverage of $0.5-4.6\,\mu$m and a redshift window of $1 < z < 2$. We find that a near-IR detection gives access to a larger, and possibly different, set of clumps within galaxies, with those also detected in UV making up only $28\%$. Whereas, $85\%$ of the UV clumps are found to have a near-IR counterpart. These near-IR clumps closely follow the UVJ classification of their respective host galaxies, with these hosts mainly populating the star-forming regime besides a fraction of them ($16\%$) that can be considered quiescent. The mass of the detected clumps are found to be within the range of $10^{7.5-9.5}\,\rm M_{\odot}$, therefore expected to drive gas into galaxy cores through tidal torques. The clump stellar mass function is found to have a slope of $-1.50 \pm 0.14$, indicating a hierarchical nature similar to that of star-forming regions in the local Universe. Finally, we observe a radial gradient of increasing clump mass towards the centre of galaxies. 
\end{abstract}




\section{Introduction} \label{sec:intro}

The observed clumpiness of galaxies, especially at $z > 1$, has been a subject of investigation over the last couple of decades. Efforts have primarily exploited the deep and high resolution imaging capacities of \emph{HST} \citep{conselice04,elmegreen05,elmegreen08,forster11, wuyts12, guo15,guo18,shibuya16, soto17, huertas-company20}. It is now widely accepted that `clumps' form as a result of gravitational instabilities in gas-rich turbulent disks \citep[e.g.,][]{immeli04, bournaud09, dekel09}. Based on rest-frame UV flux and emission-line imaging of clumps, their specific star-formation rates (sSFR) can be higher by a few factors in comparison to the rest of the host galaxy \citep{hemmati14, mieda16, fisher17, guo18, huertas-company20}. This has triggered an intense discussion regarding the significance of the clumps in galaxy evolution.

Multiple observational studies have attempted to robustly measure clump properties like their stellar mass, along with age and dust attenuation \citep{forster11, wuyts12, guo12, guo15, guo18, zanella19, huertas-company20, rujopakarn23}. The aim has been to compare these values and their gradients between two opposing theoretical models about their eventual fate. One heavily suggesting that clumps are short-lived structures ($< 100\,$Myr) and do not affect the stellar structure of galaxies besides simply increasing the net amount of stars through star-formation. They are rapidly disrupted through powerful outflows, still providing gravitational torque to direct gas towards the core of galaxies \citep{murray10, hopkins12, hopkins14, buck17, oklopvic17}. The other group of models indicate that clumps are actually massive enough to survive over tens of dynamical timescales ($\sim 400 - 1000\,$ Myr) and thereby migrating inwards to conserve angular momentum and directly contributing to the central bulge of galaxies \citep{elmegreen08, ceverino10, ceverino12, bournaud14, mandelker14, mandelker17}. One should hence expect to observe radial gradients in clump properties within galaxies.

Two of the key parameters needed to distinguish between these scenarios are the stellar mass and star-formation rate (SFR). The interpretation and possible resolution of the fate of the clumps can be linked to the interdependence of these two parameters. Disruption of clumps is related to whether the SFR density is high enough to overcome the mass dependent gravitational binding energy of these structures. Simulations have especially been making predictions regarding this connection. \cite{moody14} indicates that clumps with masses $\leq 5\,\%$ of the galaxy mass will be affected by feedback. Multiple works \citep[e.g.,][]{mandelker14, mandelker17, dekel22} however suggest that a significant fraction of clumps more massive than $\sim 10^{8.5}\,\rm M_{\odot}$ would be able to survive an inward migration due to their gravitational binding energy being sufficient to withstand the effects of feedback. Therefore, measurements of these two parameters through observations are critical. 

Detecting clumps in rest-frame UV, multiple works have concluded that $\sim 10\,\%$ of the total SFR (based on the fraction of UV light) of galaxies is contained in these structures. However, including the effects of dust attenuation complicates matters. The estimation of the stellar mass also has many challenges. The rest-frame UV selection, heavily affected by dust and stellar age, can possibly bias conclusions drawn about radial gradients observed in multiple works \citep[e.g.,][]{guo12, guo18, huertas-company20}. Opting for a rest-frame optical or near-IR selection has not been possible with \emph{HST} due to the resolution of the WFC3 instrument being insufficient to resolve the $\sim 1\,\rm kpc$ sizes of clumps. Furthermore, this size itself could just be an upper limit with a large population of clumps being smaller and less massive but beyond the detection capacities \citep{tamburello15, dessauges17,faure21}.    

With the advent of the NIRCam instrument on \emph{JWST}, we can mitigate a few of these issues. The high resolution access to the stellar-mass traced by near-IR light provides an indicator of the galaxy clumpiness that is minimally affected by dust attenuation. Although whether these wavelengths would show evidence of clumps was previously in question \citep[e.g.,][]{wuyts12, cibinel15}, it was recently shown that galaxies do show these features across rest-frame optical and near-IR wavelengths \citep{kalita24, kalita24c}. Furthermore, they are found to also be related to the overall stellar morphology of the host galaxy showing a negative correlation with the bulge dominance.  

Using the near-IR allows us to also quantify any evolution of detection limits based on stellar mass across our sample as well as over the surface of galaxies. Furthermore, having access to the stellar light distribution limits the introduction of other merging companion galaxies into the sample which might not be as apparent in shorter wavelengths and removes contamination from misidentifying galaxy bulges as clumps. This is because both low mass UV clumps and much more massive bulges as well as compact galaxies may appear similarly bright in rest-frame UV due to varying levels of dust attenuation, although they have extremely different stellar mass. These aspects are difficult to address if a study is limited to rest-frame UV. However, we still investigate the overlap between the rest-frame UV selection that has been widely carried out and our JWST/NIRCam based near-IR selection. We then carry out an 8-band spectral energy distribution (SED) model fitting using 2 HST/ACS and 6 JWST/NIRCam wideband filters with an uninterrupted (observed frame) wavelength coverage across $0.5-4.6\,\mu$m to mainly derive the stellar mass of the detected clumps. 

In this paper, Sec.~\ref{sec:sample_select} and \ref{sec:methodology} deal with the sample selection and analysis. The results are presented in Sec.~\ref{sec:results}, followed by a discussion in Sec.~\ref{sec:discussion}. The work concludes with Sec~\ref{sec:conclusion}. Throughout, we adopt a concordance $\Lambda$CDM cosmology, characterized by  $\Omega_{m}=0.3$, $\Omega_{\Lambda}=0.7$, and $\rm H_{0}=70$ km s$^{-1}\rm Mpc^{-1}$. Magnitudes and colors are on the AB scale. All images are oriented such that north is up and east is left.

\section{Sample selection} \label{sec:sample_select}


Following \cite{kalita24}, K24 from here onwards, our study targets the redshift range of $1.0 < z < 2.0$. For uninterrupted coverage of rest-frame optical and near-IR, we use the same dataset from  the Cosmic Evolution and Epoch of Reionization Survey (CEERS1; ERS 1345, PI: S Finkelstein), a JWST Early Release Science programs. CEERS1 involved NIRCam Wide-band imaging of a section of the Extended Goth Strip Hubble Space Telescope (EGS-HST) field. The reduced images \citep[made available by the CEERS collaboration;][]{bagley23} span a large wavelength window from $1\,\mu$m to $5\,\mu$m, covered by six filters (with average $5\,\sigma$ depths): F115W (29.1 mag), F150W (29.0 mag), F200W (29.2 mag), F277W (29.2 mag), F356W (29.2 mag) and F444W (28.6 mag). 

As discussed in \cite{guo15}, UV-detection of star-forming clumps is aimed at observing the rest-frame wavelength range of $2000\,$\r{A}$\,-\, 2800\,$\r{A}. The appropriate band at $1.0 < z < 2.0$ is therefore the HST/ACS F606W filter, which we access from the publicly available HST data products version 1.9, available as part of the CEERS program. The relevant mosaics were created from a combination of HST programs 10134, 12063, 12099, 12167, 12177, 12547, 13063, and 13792. This provides us with the necessary F606W filter, along with the flux in the F814W band to bridge the spectral coverage up to the F115W JWST/NIRCam filter. All images used across this study is PSF-matched to the F444W image specifications (with a point-spread function FWHM $\sim 0.14^{\prime\prime}$) using a Gaussian kernel since it has the lowest resolution\footnote{It should be noted that the F444W filter has a similar resolution to that of F814W. Therefore the degree of resolution decrease for the HST/ACS bands is minimal.}. 

The use of the Gaussian kernel does have the limitation of the JWST/NIRCam PSFs not being fully representable using a Gaussian profile. However, we address this limitation in our matching procedure. First, we fit a 2D Gaussian to each of the PSFs, including that of the reference F444W filter. The initial guess for the convolution Gaussian kernels for the three shorter wavelength filters (F115W, F150W and F356W) is based on the quadrature difference between the $\sigma$ values obtained from the respective 2D Gaussian fits and that for the F444W filter. We then convolve the PSFs with the respective (initial guess) kernels. The 2D Gaussian fitting is repeated to get updated $\sigma$ values for each filter. However, we find the new $\sigma \neq \sigma_{\rm F444W}$, with offsets $>5\%$ of the total PSF flux. So we iteratively adjust convolution kernel, repeating the process until $\sigma$ of each filter converges to $\sigma_{\rm F444W}$ within $<1\%$ uncertainty. Consequently, our PSF matching procedure depends less on PSF-Gaussian similarity and more on the azimuthal self-similarity of PSF profiles across filters.



As in K24, the sample selection makes use of the catalogue for the EGS-HST field \citep[][ S17]{stefanon17} created with an extremely large wavelength coverage ($0.4 - 8.0\,\mu$m). We limit ourselves to galaxies above the stellar mass completeness limit of $\rm log (M_{*}/M_{\odot}) = 9.5$. This is decided upon based on the $90\%$ completeness limit in S17 which varies from $log (M_{*}/M_{\odot}) = 9.0-9.5$ at $z = 1-2$, ensuring a completely unbiased sample. Using the results of the bulge-disk decomposition carried out in rest-frame near-IR (F444W, in K24), we ensure that all galaxies in our sample have a detected disk at $> 2\sigma$ significance\footnote{estimated using the uncertainty of the disk flux measurement}. Furthermore, we also limit our sample to galaxies with a disk axis ratio\footnote{The axis ratio is measured separately for the bulge and the disk in K24. We only opt for the disk ratio as it is expected to host the clumps} $> 0.3$ to ensure that we are not including edge-on galaxies which will not allow for detection of clumps. We are thus left with 418 galaxies within our sample. 

The rest-frame near-IR decomposition does not necessarily assume galaxies to perfectly follow a bulge+disk morphology. While the significance threshold set above ensures that highly disturbed galaxies are excluded, the composite model is not intended to provide a perfect fit for individual galaxies. Instead, we uniformly separate all galaxies in our study into bulge and disk components (with fixed Sérsic indices), with resulting uncertainties reflected in the flux measurements. This approach is widely used \citep[e.g.,][]{simard11, meert11, bottrell17a, bottrell17b, bottrell19} and provides a consistent metric for spatial decomposition.


\section{Methodology} \label{sec:methodology}
This paper aims to investigate the property of clumps primarily in near-IR along with UV. For the near-IR wavelength window, we used the JWST/NIRCam F356W band as it covers rest-frame wavelengths $>1\,\mu$m across our redshift range and the wavelength is high enough to minimise the effect of dust attenuation. We do not use our longest filter F444W as it is 0.6 magnitude shallower than F356W. A maximum attenuation of A$_{V} = 3$ will lead to F444W being only $\sim 0.1$ magnitude brighter, for a typical \cite{cardelli89} Milky-way extinction curve. Hence it is not enough to compensate for the shallower depth of F444W in comparison to F356W. Nevertheless, to map the extent of stellar emission of each galaxy within our sample, we use the F444W filter as it still is the least susceptible to variations in dust attenuation. Without the need of detecting substructures, it performs the best in spatially deblending galaxies from neighbours. In case of the rest-frame UV, we use HST/ACS F606W to correspond to previous literature studying UV clumps \citep{wuyts12, guo15, guo18, sattari23}, which target the rest-frame wavelength range of $2000\,$\r{A}$ -\,2800\,$\r{A}.   

\subsection{Wavelet decomposition} \label{sec:wavelet_decomp}

\begin{figure*} 
    \centering
    \includegraphics[width=0.9\textwidth]{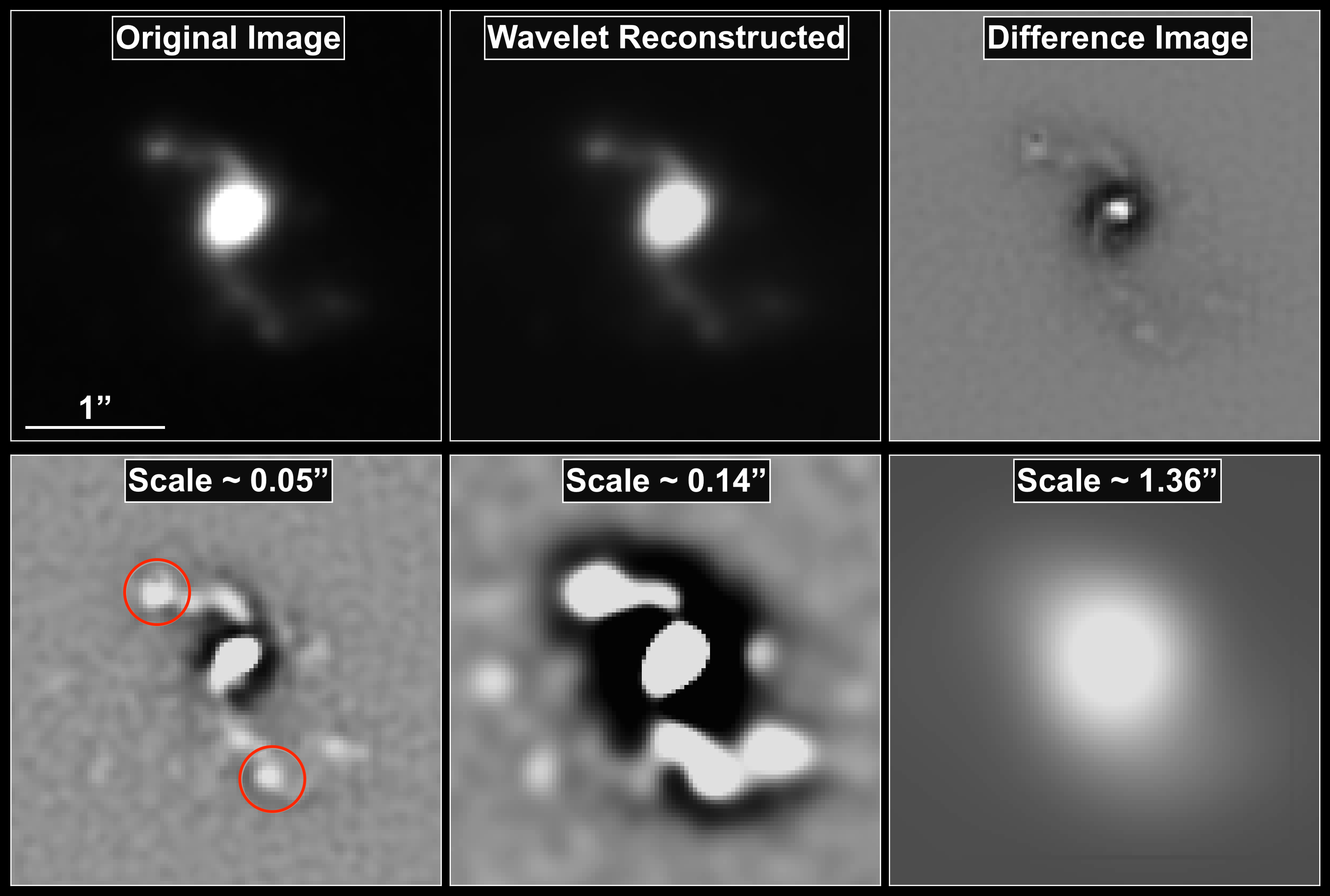}
    \caption{An illustration of the wavelet decomposition process using Starlet transform on a sample galaxy (S17 ID 22922). (Top row) The first image is the original F356W image, followed by the reconstructed image created by co-adding all wavelet scales. The final is the difference between the first two, showing the efficacy of the wavelet decomposition. The flux unaccounted for makes up this residual and is found to be $3\%$ of the total flux in the orginal image. (Bottom row) Three out of the seven wavelet scales that the original image was decomposed into, which spans a total range of $\sim 0.01^{\prime\prime} - 1.36^{\prime\prime}$. On the image of scale $\sim 0.05^{\prime\prime}$, we mark in red the location of the clumps later detected using the algorithm discussed in Sec.~\ref{sec:clump_detection}.}
    \label{fig:wavelet_decomp}
\end{figure*}

The images first require a subtraction of larger scale structures from a given galaxy image in order to enhance the small-scale clump detectability. This process was done in K24 through the convolution of the original image with a circular Gaussian kernel of $\sigma = 4\,\rm pixels\,(0.12^{\prime\prime})$, hence creating an effective `background', and then subtracting it away to leave behind a `contrast' image. The same method has been widely used in previous studies \citep{conselice03, guo15, guo18, calabro19}. However, this method has a few limitations:

\begin{itemize}[leftmargin=*, noitemsep, topsep=0pt]
    \item The smoothed image is a convolution of the original version with a Gaussian. The contrast image hence derived has the issue of self-subtraction, especially for the brightest clumps and galaxy cores that usually feature strong peripheral negative peaks. 
    \item The use of the Gaussian kernel for smoothing relies on the assumption of symmetry of the structures, which is not always appropriate.   
    \item The scale length, most sensitive for the contrast image, is difficult to determine. The result of the subtraction effectively allows detection of regions varying at scales determined by the difference of the original PSF of the image and that after smoothing.   
\end{itemize}

A way to overcome these is to use Fourier decomposition of the image and apply a galaxy specific low-pass filter in the frequency domain. This was shown to be effective in the background subtraction of rest-frame UV images \citep[from HST/ACS;][]{sattari23}. However, we found that using a Fourier transform on the visibly more complex JWST/NIRCam images comes with a drawback. Due to more detectable `edges' or discontinuities in galaxy morphologies, we face the well-known issue of Gibb's ringing. This can create artificial clumps near bright regions of galaxies.      

The workaround is to use a different basis function in the (wavelet) transform, which is more attuned to the goal of modelling galaxy morphology. We implement the Isotropic Undecimated Wavelet Transform (also called starlet wavelet transform), which is well adapted for astronomical data \citep[][for a review]{starck07}. We show the use of this method for the F356W image of one of the galaxies within our sample in Fig.~\ref{fig:wavelet_decomp}. The wavelet transform decomposes the original image into seven images of increasing scales\footnote{The scales are determined to be approximately $0.01^{\prime\prime}$, $0.05^{\prime\prime}$, $0.08^{\prime\prime}$, $0.14^{\prime\prime}$, $0.27^{\prime\prime}$, $0.55^{\prime\prime}$, $1.36^{\prime\prime}$. These values are effectively the FWHM of a Gaussian fit applied separately on the seven images from the wavelet decomposition of a single point source image}, three of which are shown in the figure for illustration.  

To check the efficacy of the transform, we reconstruct the galaxy image using all the wavelet scales and subtract this from the original image. Based on the difference image, we conclude that the net flux unaccounted for in the reconstruction is $3\%$, which we deem to be satisfactory. Nevertheless to account for this, during the clump detection procedure we subtract the larger wavelet scales from the original image rather than simply adding the smaller scales. This conserves the remaining $3\%$ of the flux. We also find it ideal to subtract the scales equal to and greater than $0.14^{\prime\prime}$ for detecting the clumps in our sample galaxies. This is we find to be appropriate since the $0.14^{\prime\prime}$ scale is the point-spread-function FWHM for the images we use (Sec.~\ref{sec:sample_select}). A further reasoning behind the use of this specific value has been discussed in detail in the Appendix. However, it should be noted that this places a spatial detection limit of $0.5-1.2\,$kpc for our redshift window, with our method unable to detect smaller structures. 

\begin{figure*} 
    \centering
    \includegraphics[width=\textwidth]{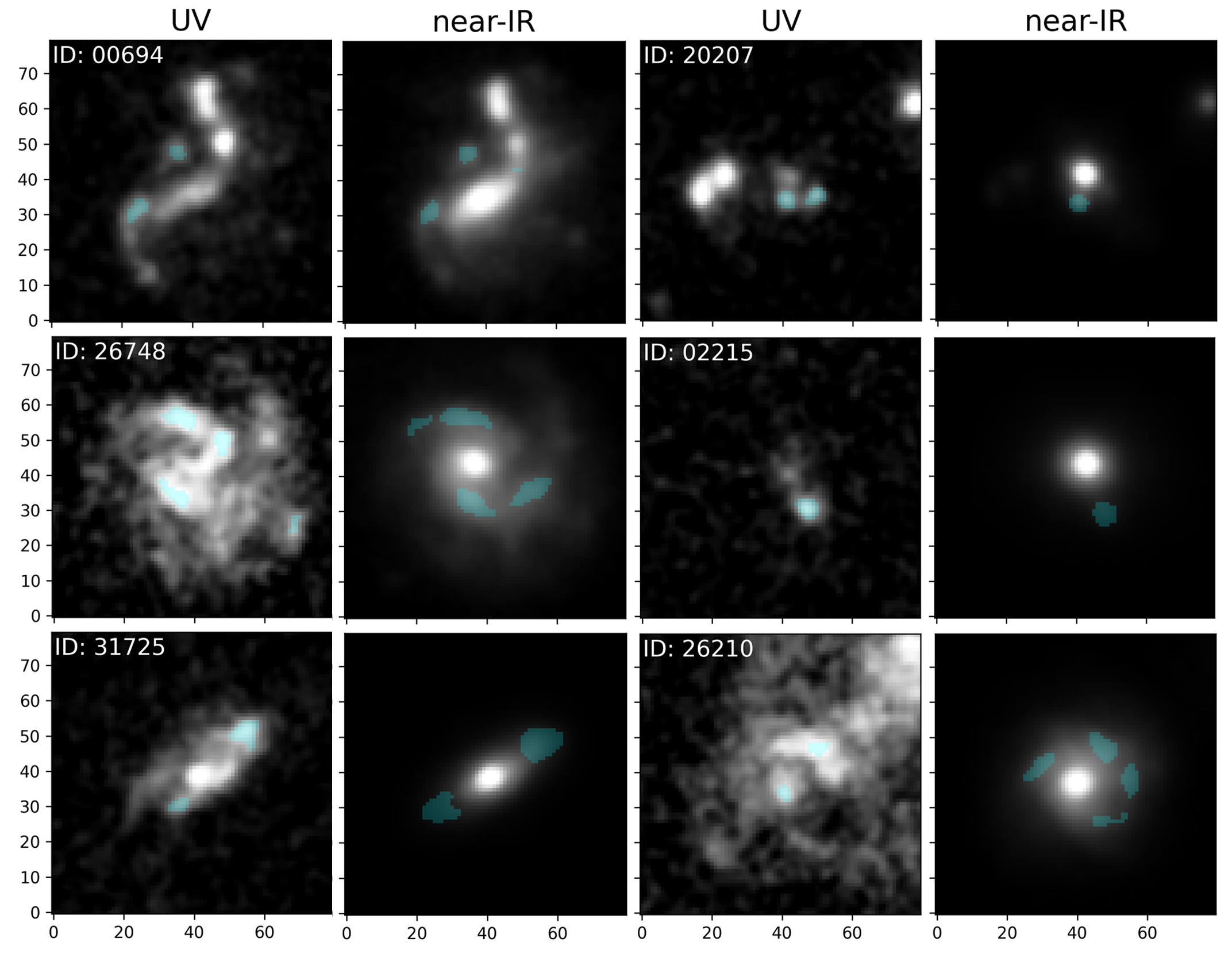}
    \caption{A compilation of the rest-frame UV (F606W) and near-IR (F356W) images of galaxies with detected clumps, highlighted with a blue color scheme. It should be noted that the clumps may not always be clearly detected by eye on these original images, for which one would rather require the respective contrast maps. Furthermore, the detection of clumps are restricted to a $7\,\sigma$ region of the corresponding F444W image and therefore structures outside this region would not be selected, as can be observed in a few of the examples. The corresponding S17 catalogue IDs are provided on the top left for each source.}
    \label{fig:image_composite}
\end{figure*}

\subsection{Clump detection} \label{sec:clump_detection}

After the subtraction, we implement a source detection on the resulting contrast image using the python package PHOTUTILS \citep{photutils}. The target region within which the clumps will be searched for is determined through a $7\,\sigma$ source detection in the F444W image. The value is chosen such that the resulting region is found to be most consistent with the $90\%$ flux radius of the disk, derived from the bulge-disk model previously fit to the F444W image. This strict threshold, which is used with a deblending contrast parameter of $\sim 7.5\,$ magnitude, ensures exclusion of any neighbouring galaxies in the F444W image within the source region. This significantly reduces the possibility of including interlopers or neighbouring galaxies interacting with the primary source. Examples of this can be observed in Fig.~\ref{fig:image_composite} for IDs 694 and 20207, where massive structures near the primarily galaxy have not been identified as clumps. This is aimed at selecting clumps strictly within the extent of the galaxy stellar disk, likely also removing merger companions.

To ensure inclusion of all galaxy sub-regions, we visually inspect each source segmentation map for the 418 galaxies in our sample. We also examine maps that use $3\,\sigma$ and $5\,\sigma$ thresholds, which are often found to include potential companions. We also try using a search region equal to the $90\%$ flux radius of the disk, derived from the bulge-disk model previously fit to the F444W image. This results in only $\sim 2\%$ change in total clump count, due to the $7\sigma$ threshold being set close to the $90\%$ flux radius. We also consider using the deeper F356W image to define the search region, but this approach often splits galaxies into sub-components because of the increased depth and prominence of substructures, potentially excluding some galaxy regions. Although our detection procedure aims to minimize exclusion of genuine clumps within the galaxy disk, this possibility cannot be fully dismissed and should be noted as a caveat.

The clump detection threshold is set to $4.5\,\sigma$ for the F356W residual image (lower than the $5\,\sigma$ threshold set for F200W in K24, based on the relative depth in F356W). For the F606W, which is $0.6$ magnitude shallower, we use a threshold of $3.0\,\sigma$ to account for the difference in depths. This however assumes a flat spectrum ($F_{\lambda}$), which may not be accurate. But the UV selection will inevitably depend on the dust attenuation, hence one anyway cannot determine consistent detection limits as a function of stellar mass of the structures. This estimation of stellar mass limits is only possible in F356W (discussed later in Sec.~\ref{sec:mass_completeness}) making this form of detection more robust across our sample. During the clump identification, we also ensure that the central core is not selected, which appears as a clump especially in the F356W. The determination of which of the clumps is the core is done using the coordinates returned from the bulge-disk decomposition\footnote{This value is found to always be consistent with the asymmetry centre of the galaxy in F444W, thereby confirming the location of the center.} that was undertaken for all galaxies within our sample in K24 using the stellar mass tracing F444W flux distribution.

Using the detection images from F606W and F356W filters, we obtain the two sets of segmentation maps containing detected clumps that we will refer to as UV and near-IR clumps respectively for the rest of this work (Fig.~\ref{fig:image_composite}). For each clump, we asses the robustness of its detection by duplicating its flux distribution at a new random location in the original galaxy image (within the search radius) and by rotating it by a random angle. Then the image is passed through the whole process starting from the wavelet subtraction. Only when a clump is detected $> 68\,\%$ (corresponding to $1\sigma$ for a Gaussian distribution) of the 100 iterations we conduct do we keep it within our sample. This method removes $\sim 15\%$ of the original detection, providing a filtered version of clump segmentation maps. The flux within the respective regions in these maps makes up the clump flux, and the standard deviation in the distribution of flux measured over the 100 iterations combined with the image noise gives the corresponding uncertainty. These segmentation regions are then used to measure the clump flux in all remaining filters.

For the clump flux measurements the results of this work uses flux within apertures on the original images. This is mainly to ensure our detection limit estimation (Sec.~\ref{sec:mass_completeness}) can also be used in parallel to assess the significance of our results, as will be discussed at length throughout this work. We further investigate the use of a background subtraction using an average flux within an annulus set at the radial distance of each clump from the centre of their host galaxies. The subsequent results and conclusions are identical except for a small decrement of the average stellar mass estimation of clumps (by $\sim 0.1-0.2$ dex). However, we observe that the annulus method might lead to over-subtraction in many cases since at longer wavelengths, besides the detected clumps, there exists substructures which lead to an overestimation of the background. This will be discussed in detail later in Sec.~\ref{sec:mass_discussion}. 

Furthermore, the flux measurements can also be made from the contrast images. This would be a severe underestimation by about an order of magnitude since we are essentially removing the flux in larger scales which would still account for a percentage of the clump flux. The wavelet decomposition of small structures still associates some flux to the larger scales and is only efficient for detection rather than for measurement.  Nevertheless, we mention the results from the contrast image, especially in regards to radial property variation as an extreme limit across the work and also in the Appendix. We do not however determine stellar masses by this method.

\subsection{SED fitting} \label{sec:fast_sed}
After the measurement of fluxes of the clumps over an observed wavelength range of our data ($0.5-4.6\,\mu$m), we derive their physical properties through spectral energy distribution (SED) modelling. For this, we use the software FAST++\footnote{https://github.com/cschreib/fastpp}. To simplify the fitting, we allow the redshift to only vary within the $68\,\%$ range provided for the respective host galaxies in S17. We use the \cite{bruzual03} stellar population models with the \cite{chabrier03} initial mass function, along with the \cite{calzetti00} dust attenuation law. We fix the metallicity to the solar value\footnote{Since the primary objective of this work is to determine the stellar mass, this property does not vary beyond the already present uncertainties at fixed metallicity which we separately verify for a sub-sample using the Bayesian-based SED fitting code \textit{BAGPIPES}. This lack of stellar mass dependence on metallicity has also been recently shown in \cite{osborne24}.}, while allowing the dust attenuation to vary across the range A$_{V} = 0-6$. We implement a simple exponentially declining model ($\propto e^{-t/\tau}$) as the star-formation history (SFH) model. We also attempt to use the delayed-exponentially declining SFH $\propto (t/\tau^{2})\,e^{-t/\tau}$. The results from the latter, although in agreement with those from the exponentially declining model, have slightly larger uncertainties in measured parameters likely due to the increased complexity of the SFH. Therefore, we will be referring to the results from the former model throughout this paper.  


Given that our observations do not include wavelength coverage into far-IR and sub-mm where dust-reprocessed stellar light is emitted, we will abstain from using the exact measured values for star-formation within the clumps to draw any conclusions. Nevertheless, we exploit an UVJ color-color selection to determine a general star-formation classification discussed later in Sec.~\ref{sec:clump_classification}. Similarly, we also do not discuss stellar age and dust attenuation values beyond a brief mention as it can be highly uncertain without spectroscopic data. The critical $4000\,$\r{A} break that is required to robustly determine the stellar age is straddled by the F814W HST band and the F115W JWST band. The shallow depth of the former filter results in uncertainties that leads us to not rely on the measured age values.  

In addition to carrying out this SED fitting procedure for individual clumps, we also perform the SED fitting using the integrated fluxes for the whole galaxy within the region that we demarcated using a $7\,\sigma$ threshold in the F444W image (Sec.~\ref{sec:clump_detection}). This provides us with a revised host galaxy stellar mass, which will be used for the rest of this work. It should be noted that in a fraction of cases, the stellar mass is lower than reported in S17. This is mainly prevalent in galaxies at the lower mass end of our sample. We expect the underestimation since the region of selection is dependent not on the actual flux of the galaxy but rather the depth of the data. However, we argue that the stellar mass we derive from the $7\,\sigma$ threshold ensures that we only include the region where the clumps are searched. Using a radius dependent on the galaxy flux (e.g., the kron radius) as the selection region would also require a correlated clump detection limit, which would severely increase systematic biases. We then proceed to check the galaxy $90\%$ completeness limit ($= 10^{9.5}\,\rm M_{\odot}$) based on the currently defined threshold and limit our study to the sources with the revised stellar mass above it.

\subsection{Detection limits} \label{sec:mass_completeness}

The best constrained physical parameter obtained from SED fitting is the stellar mass of the clumps, compared to their stellar ages and star-formation rates. This accuracy ($\sim 0.2\,\rm dex$) can be attributed to the extensive coverage of the rest-frame optical and near-IR with the JWST/NIRCam filters. However, assessment of the stellar mass detection limits across our sample is important for this study. This estimation is done by injecting a point source\footnote{The choice of the point source is due to the FWHM being equal to the size limit set by our wavelet subtraction method (Sec.~\ref{sec:wavelet_decomp})} of varying flux ($-4.0$ to $0.0$ in log scale relative to the flux of the entire galaxy; see also K24) into the F356W images within the confines of the $7\,\sigma$ based demarcation of the extent of each galaxy. It is also ensured that these positions do not overlap with the core or any already present clumps. Only the limits for the near-IR clumps (in F356W) are discussed in the work, as the range of masses possible for a specific UV flux limit is too high ($\gtrsim 1\,\rm dex$, compared to $\sim 0.3\,\rm dex$ in near-IR\footnote{these ranges are obtained using the values of mass-to-light ratios of the set of clumps in every clumpy galaxy}) for consideration due to contribution from both star-formation and dust attenuation.  

The location of the artificial clumps are varied across the radius of the disk, thereby estimating the limits as a function of radial distance from the centre of the galaxy. Given that this procedure is being carried out using the rest-frame near-IR flux which directly traces the stellar mass, we use the ratio of the detection limit and the flux of the whole galaxy to obtain the associated limit in stellar mass. This assumes that the SED of the artificial clumps would on average be similar to that of the whole galaxy. Such a correspondence of near-IR clumps and the host galaxy is later verified in Sec.~\ref{sec:clump_classification}. We also carry out this detection limit estimation using the annulus subtraction method (Sec.~\ref{sec:clump_detection}). However, the limit in some cases are found to be negative especially within the central region of some galaxies. We conclude that underlying flux gradients at longer wavelengths, likely due to stellar substructures, can be higher than the lowest flux of a clump that can be detected at the same radial distance.

\section{Results} \label{sec:results}

\subsection{Clump detection in near-IR vs UV} \label{sec:clump_selection}

One of the main goals of this work is to understand the significance of near-IR detection of clumps in place of UV. Firstly, we find that $40\,\%$ of our sample galaxies show clumps in near-IR within our detection thresholds (with the $68\%$ completeness limit at $10^{8.1}\,\rm M_{\odot}$ for the most massive galaxies). Each of these clumps are expected to be within the stellar disk of the host galaxy as observed in the F444W filter. Therefore the inclusion of merging neighbours are severely limited.  

Such a systematic detection is not possible using UV where the flux is dependent on the star-formation as well as dust obscuration. Hence the UV detection limit is found to vary over a magnitude in stellar mass. The UV flux threshold we set assuming a flat spectrum (Sec.~\ref{sec:clump_detection}) results in a detection of clumps in $25\,\%$ of our sample. Moreover, $82\,\%$ of these UV clumps have a near-IR counterpart, suggesting most of these objects are also detected in near-IR. However, converse is not true with only $28\,\%$ of near-IR clumps having a spatial counterpart in UV. Taking the above mentioned results into consideration, we conclude that using the rest-frame near-IR allows for a more standardized detection of clumps in galaxies in terms of their stellar mass.

\subsection{Clump luminosity function}
\begin{figure} 
    \centering
    \includegraphics[width=0.5\textwidth]{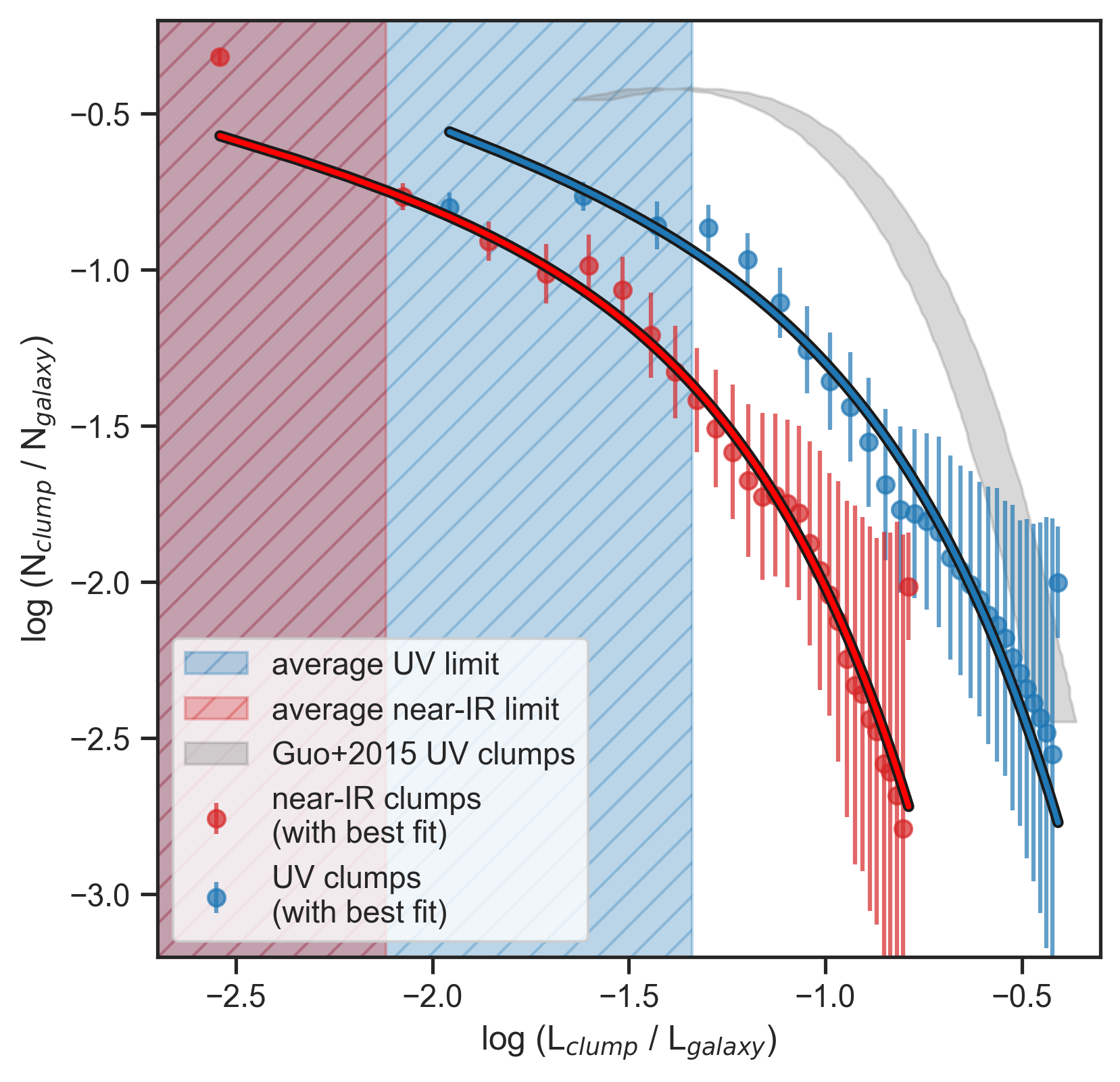}
    \caption{The fractional luminosity function, which shows the normalised clump number count as a function of the luminosity ratio between each clump and the host galaxy at the same observed wavelength. The near-IR clumps are in red and the UV clumps are shown in blue. The corresponding best fit Schechter Functions are provided for each set. Also displayed are the average $68\%$ detection limits for both rest-frame near-IR and UV. For comparison, the distribution from a previous HST study \citep{guo15} is also provided.}
    \label{fig:clump_lum_fucntion}
\end{figure}

The detection of clumps in rest-frame UV vs. near-IR can be further explored using a clump fractional luminosity function, which normalizes clump count as a function of the clump-to-host luminosity ratio (fractional luminosity). This function for both UV and near-IR clumps is shown in Fig.~\ref{fig:clump_lum_fucntion}. The clear offset in fractional luminosity indicates a difference in contrast with the host and the contribution from the near-IR bright bulge within the galaxy flux. However, the similar evolution of the distributions suggests an intrinsic similarity between the two populations.

The detection of clumps in rest-frame UV allows us to also compare our analysis to previous literature. The largest of these is the \cite{guo15} study of 3239 galaxies across a redshift range of $0.5-3.0$, with HST/ACS data. As discussed in Sec.~\ref{sec:methodology}, over the redshift range of $1 < z < 2$ the F606W filter was used for clump detection allowing a direct comparison to our study. However, some key differences still exist. The final detection was done with at $2\sigma$, whereas we implement a $3\sigma$ threshold. Furthermore, no additional detection checks were conducted for each clump, as a result of which we reject $15\%$ of our clumps. The background estimation is also different as their study implements an estimation using a region of $6-10$ pixels around the location of the clump. We find this to approach the disk $r_{e}$ and is found to be lower the clump flux by a factor of $\sim 0.8-0.9$ in comparison to the flux measurements without any background subtraction. Finally, since there was no access to resolved rest-frame near-IR data at these redshifts before JWST, the clumps which actually are the galaxy bulges are not removed.

For the comparison, we plot the same clump fractional luminosity function in \cite{guo15} for $1 < z < 2$ and galaxy stellar mass range of $10^{9.8-10.6}\,\rm M_{\odot}$. Although our UV clumps overlap with the luminous end of the distribution, we do not see an agreement. We attribute it to our conservative detection algorithm that likely lowers the normalisation of the luminosity function. Our method includes the removal of bulges as well as fainter clumps during our robustness checks. Rejection of clumps beyond the detection region will also be a possible contributor to this offset, especially since less massive clumps are likely to be at larger radii (Sec.~\ref{sec:spatial_distribution}).

We emphasize that differences in the detection process inevitably lead to inconsistencies in the resulting clump populations. This is primarily because no single detection process is ideal, and each comes with its own limitations. A recent study by \cite{kalita24c} presents a more detailed analysis without clump filtering. They adopt a $3-4\,\sigma$ detection threshold and ensure the galaxy, along with its clumps, is well-fitted using a composite morphological model. However, even in that study, some residual structures or clumps remain unincorporated.

\subsection{Mass of clumps} \label{sec:clump_mass}

\begin{figure} 
    \centering
    \includegraphics[width=0.48\textwidth]{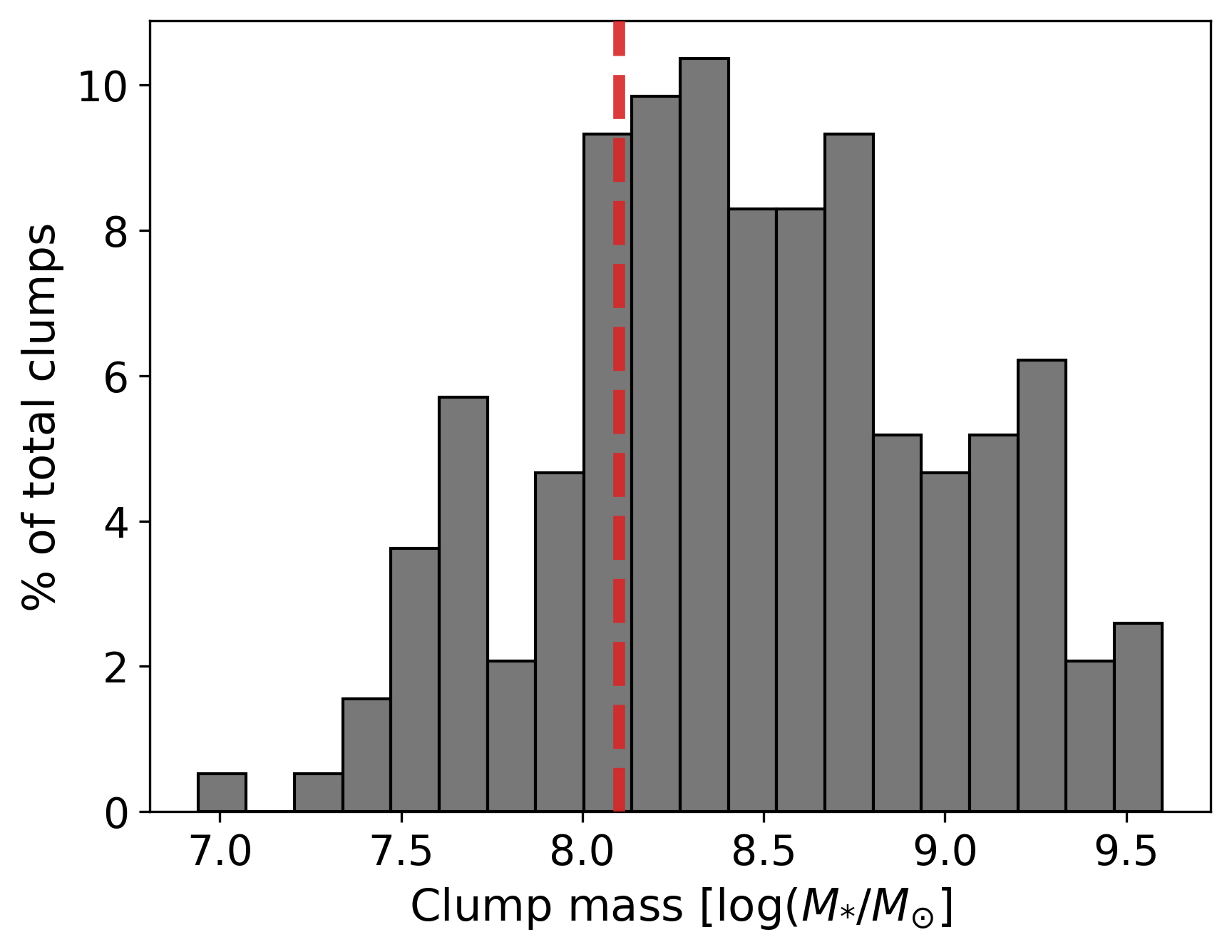}
    \includegraphics[width=0.46\textwidth]{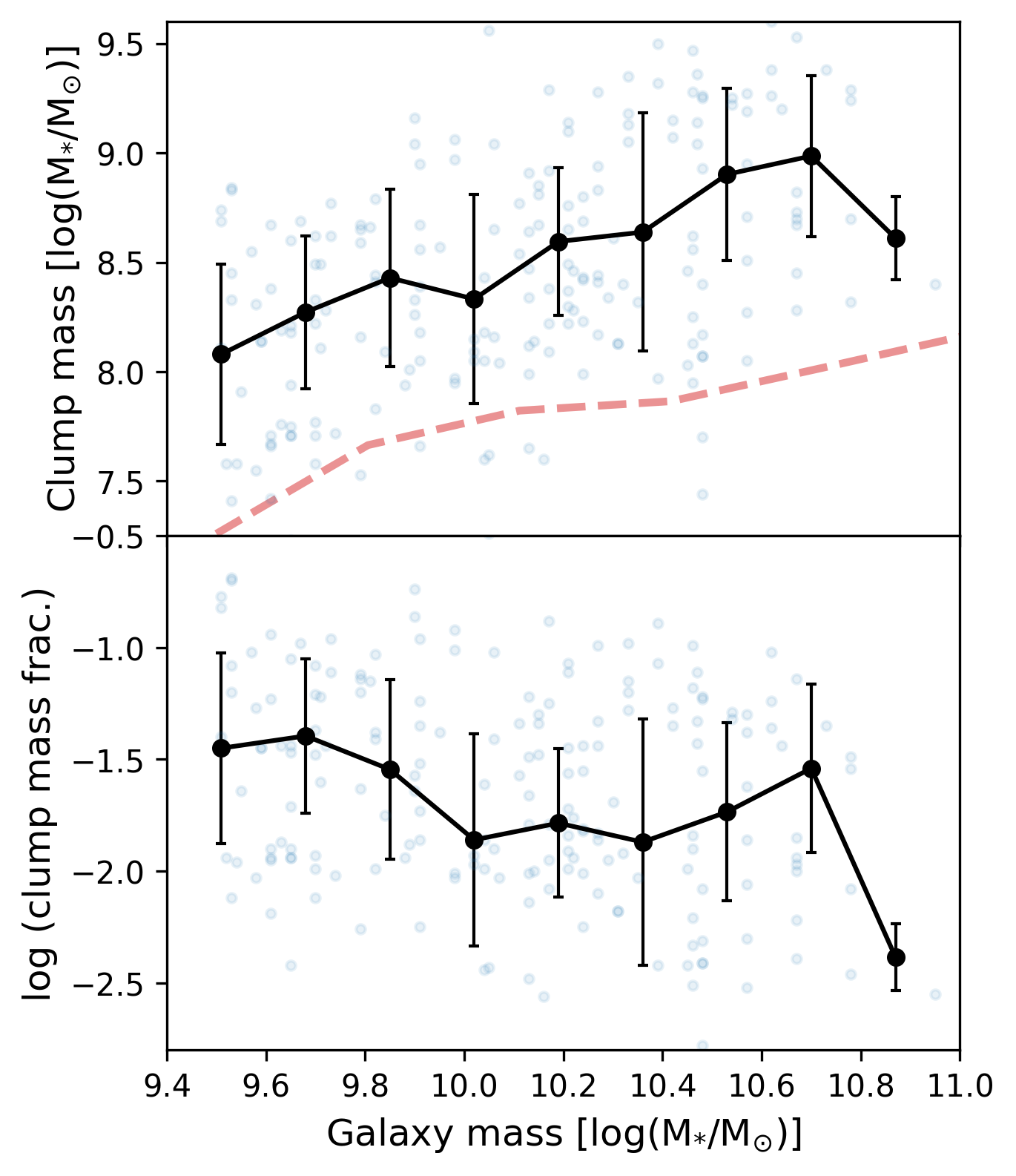}
    \caption{(Top) The distribution of clump mass across our sample, with the red-dashed line indicating the detection limit for the most massive galaxies in our sample. (Middle) The stellar mass of clumps, in y-axis, as a function of the host galaxy stellar mass. The black points denote the average trend with the error bars representative of the dispersion. The red-dashed line indicates the completeness limit of our detection algorithm across the host galaxy mass distribution. (Bottom) The y-axis provides ratio of the stellar mass of clumps and the host galaxy mass. In each of the aforementioned plots, the data points correspond to individual clumps, with more than one of these possibly belonging to the same galaxy.}
    \label{fig:clump_mass_plot}
\end{figure} 

\begin{figure} 
    \centering
    \includegraphics[width=0.5\textwidth]{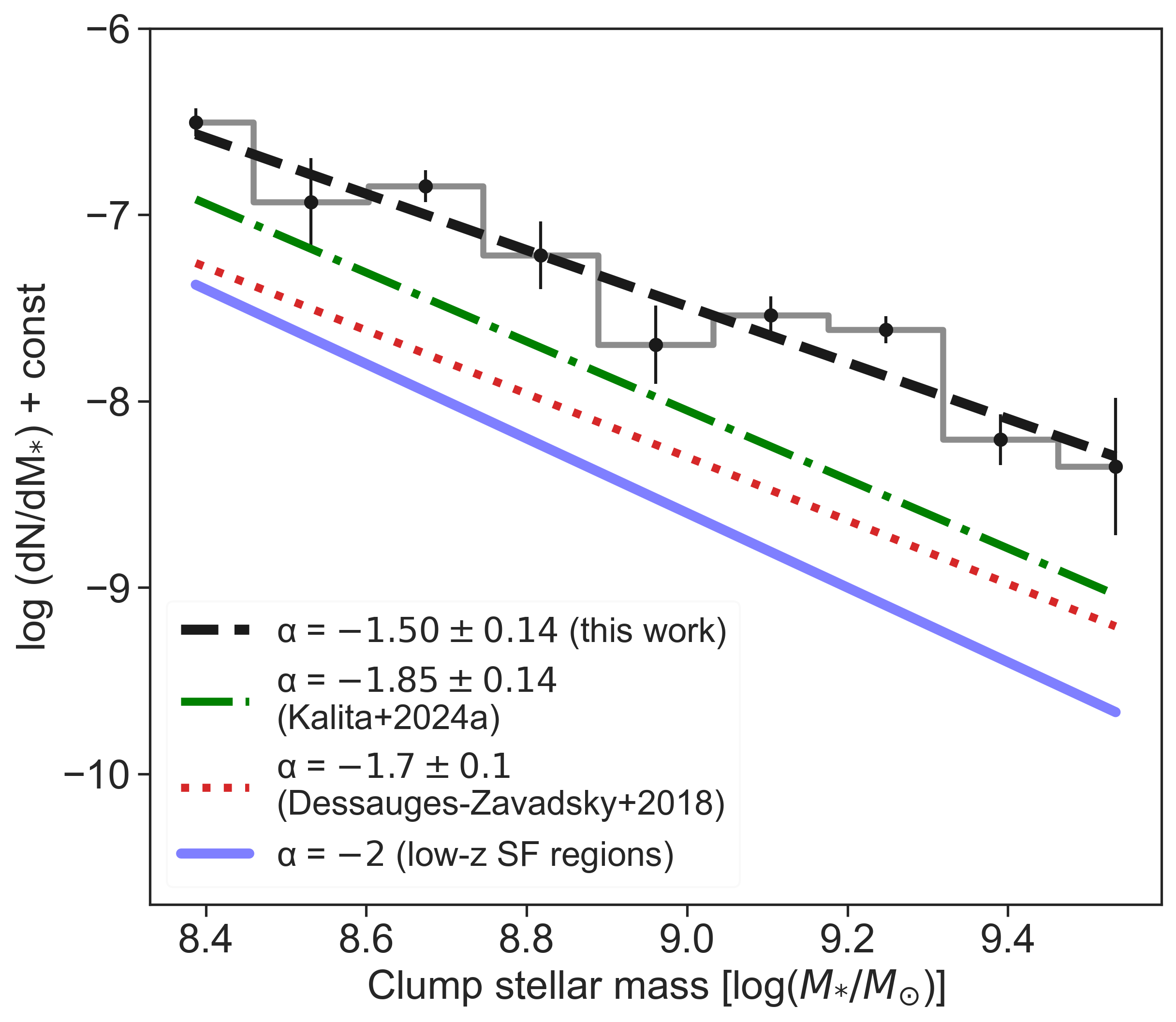}
    \caption{The clump stellar mass function in our sample. This is compared to the theoretical prediction for hierarchical star-forming regions \citep[$\alpha = -2$][]{elmegreen06} as well as the function derived from clumps in lensed galaxies \citep{dessauges18}. A final comparison to the model-based clump analysis \citep{kalita24c} in 32 massive galaxies ($>10^{10.5}\,\rm M_{\odot}$) is also provided.  A constant has been added to the power law for a visual comparison of the slopes and does not hold any physical relevance.}
    \label{fig:clump_smf}
\end{figure}

In this section about the stellar mass, we limit our discussion to near-IR clumps (due to the reasons specified in Sec.~\ref{sec:clump_selection}). We find the distribution of clump mass to be between the stellar mass range of $10^{7.5-9.5}\,\rm M_{\odot}$ (Fig.~\ref{fig:clump_mass_plot}, top) with the average being $\sim 10^{8.5}\,\rm M_{\odot}$. Meanwhile the $68\%$ completeness limit is found to be at $10^{8.1}\,\rm M_{\odot}$, with our method not suitable for detecting clumps with lower masses especially in massive ($> 10^{10.5}\,\rm M_{\odot}$) galaxies. We see a drop in the number of clumps below this limit in Fig.~\ref{fig:clump_mass_plot}, which we regard as a systematic effect rather than a real physical characteristic. Based on these measurements, we also model the clump stellar mass function (cSMF). We fit a power law (log($dN/dM$) = $\alpha \times $ log $M$ + log $\rm (const)$) to the clump number counts above the completeness limit as a function of stellar mass and find $\alpha = -1.50 \pm 0.14$ (Fig.~\ref{fig:clump_smf}). The normalisation depends on detection thresholds and the input parameters of the SED fitting procedure \citep{dessauges18}, making direct comparisons between different studies challenging. Hence the constant that has been added is simply for the ease of visual comparison in Fig.~\ref{fig:clump_smf} and does not hold any scientific relevance. 

We show the distribution of clump mass in Fig.~\ref{fig:clump_mass_plot} (middle), in reference to the corresponding host galaxy mass. We observe a slight trend of the clumps progressively growing more massive with increasing host galaxy mass. However, the scatter is high and the increase can be partially compensated by the variation in the completeness limit we provide in red (as determined in Sec.~\ref{sec:mass_completeness}). Hence it is likely that we are not detecting clumps less massive than $10^{8.1}\,\rm M_{\odot}$ in massive galaxies ($> 10^{10.5}\,\rm M_{\odot}$). Detecting them would inevitably increase the scatter at the high galaxy mass end. Nevertheless, we still can observe that the most massive clumps ($\gtrsim 10^{8.5}\,\rm M_{\odot}$) are preferentially located in massive galaxies.

If we inspect the ratio of the clump mass and the host galaxy mass (Fig.~\ref{fig:clump_mass_plot}, bottom), we find a marginal decrease of about $\sim 0.5\,\rm dex$. However, the change is not sufficiently significant with respect to the scatter of the data. Finally, we detect a decrease in the mass and fraction for clumps at the highest stellar mass bin of $10^{11}\,\rm M_{\odot}$. This is especially interesting because the galaxies sampled in this bin are mostly in the UVJ-based quiescent regime.

\begin{figure*} 
    \centering
    \includegraphics[width=\textwidth]{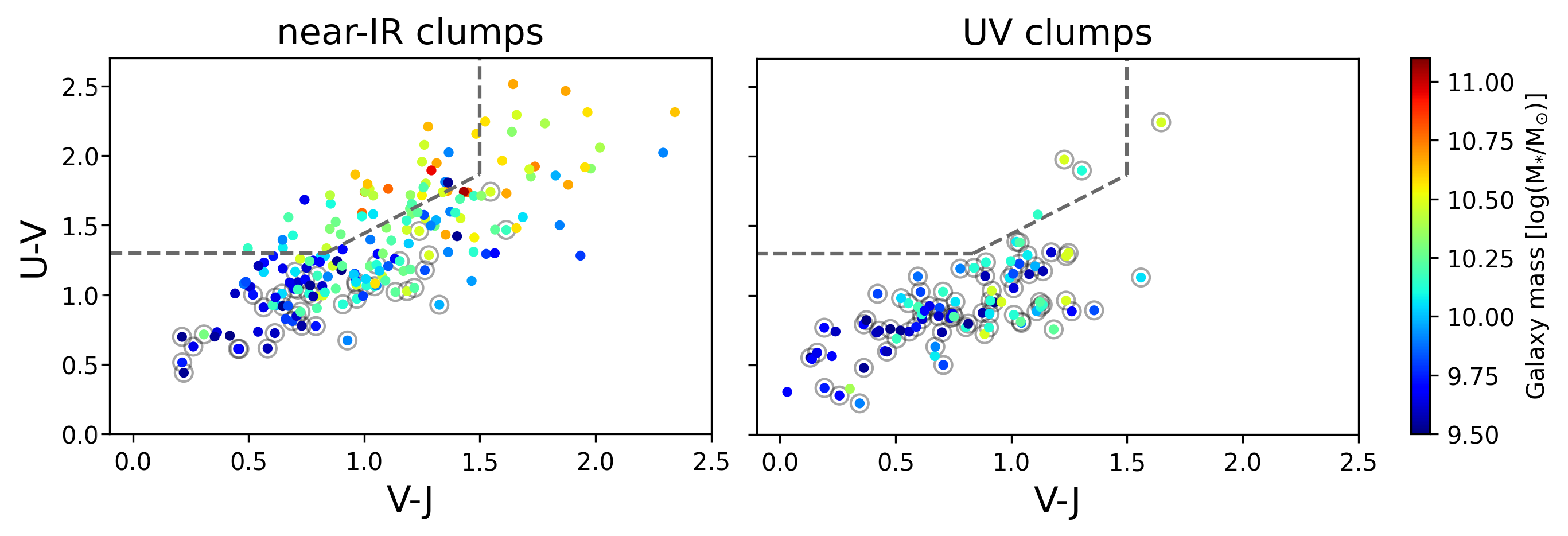}
    \caption{(Left) The UVJ plot (all values in AB magnitudes) for individual near-IR detected clumps, with grey circles indicating the ones with corresponding UV detection. The color of the points refers to the host galaxy mass. (Right) The same plot for the UV clumps, with those having corresponding near-IR detection featuring additional grey circles. In both the plots, the top-left region demarcated by the dashed lines indicate the quiescent regime, while the rest can be regarded as star-forming.}
    \label{fig:uvj_clumps}
\end{figure*}
\begin{figure*} 
    \centering
    \includegraphics[width=\textwidth]{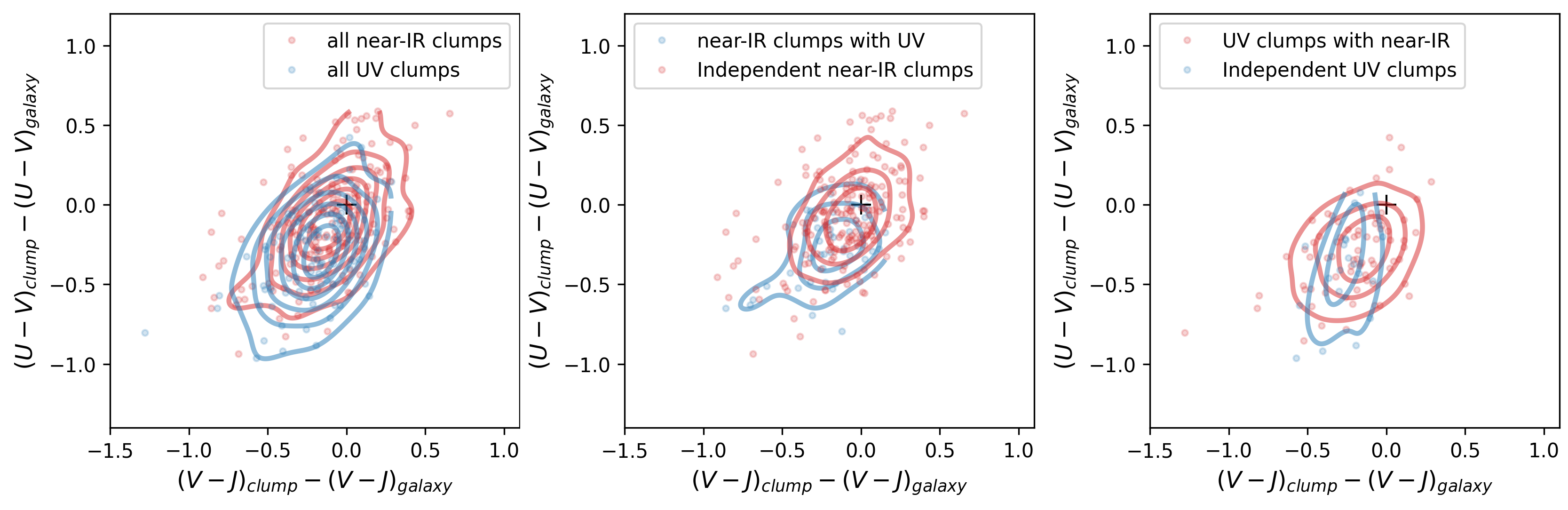}
    \caption{(Left) The U-V and V-J (all values in AB magnitudes) color difference of all near-IR and UV clumps, with respect to their respective host galaxies. The black `+' sign marks the 0 value, with point lying to the bottom left being less dust obscured and to the top right being more dusty. Meanwhile, values from bottom right to top left have decreasing specific star-formation, with the 0 value suggesting similar value as that of the whole galaxy. (Middle) The same plot, but only for UV clumps, divided into two groups: with and without corresponding near-IR detection. (Right) The distribution for near-IR clumps, with and without UV detection.}
    \label{fig:uvj_clumps_galaxy_diff}
\end{figure*}

\subsection{Color-color classification} \label{sec:clump_classification}

We implement the SED fitting results to classify each clump based on its U-V and V-J colors (Fig.~\ref{fig:uvj_clumps}). This is done using the fitted SEDs for each object that are within $\Delta \chi^2 = 1$, based on which the rest-frame U, V and J band magnitudes are derived. Throughout, we use the filters prescribed in \cite{whitaker11}, while also following their UVJ classification criteria for quiescent (top-left of the figures demarcated by the dashed lines) and star-forming galaxies. Within the star-formation regime, progression from the bottom-left to top-right translates to an increase in dust attenuation. While a trend from bottom-right to top-left indicates a decrease in specific star-formation rate which finally leads to a transition into the quiescent galaxy regime. To ensure quality-control, only those measurements are plotted for which the SED fitting returns a reduced-$\chi^{2} = 1\pm 1$.

The near-IR clumps show that they primarily lie in the star-forming regime of the UVJ diagram. Although, a small fraction of them do occur in the quiescent region, indicating that one can have clumps with minimal star-formation. The UV clumps however populate only the low dust-attenuation segment of the star-forming region, which is expected since clumps being detected in UV implicitly suggests high star-formation devoid of dust. Focusing on the host galaxy stellar mass of the clumps, we observe that UV clumps primarily appear in galaxies below a mass of $\sim 10^{10.5}\,M_{\odot}$, near-IR clumps are featured in the whole span of our galaxy mass range. For both the sets of clumps, there is a clear gradient of increasing host galaxy mass towards the top right hand-side of the UVJ diagram. This suggests a gradual increase in dust content and/or reduction in star-formation in clumps as we progress towards more massive galaxies. 

Finally, we place the UVJ colors of clumps in direct reference to those for their respective host galaxies in Fig.~\ref{fig:uvj_clumps_galaxy_diff}. The near-IR clumps mainly follow the host galaxy properties albeit with an ubiquitous scatter of $\sim 0.4$. This suggests that there can be variations in specific star-formation and dust-attenuation within the surface of individual galaxies. However, the scatter along the axis of changing dust-attenuation is observed to be higher than that for specific star-formation. Nevertheless, this scatter is well below the total extent of the UVJ positions of both the complete galaxy or the clump sample. Hence, Fig.~\ref{fig:uvj_clumps_galaxy_diff} suggests that the clump properties in general follow the host galaxy. Meanwhile, the UV clump distribution concentrates towards the low attenuation regime, which corroborates Fig.~\ref{fig:uvj_clumps} 

We do note that given our results are based on the clump flux measurements on the original images rather than the contrast maps, the correlation between the clumps and the host galaxies may simply indicate a systematic effect from galaxy flux contribution. We therefore redo the whole analysis with the contrast maps where the galaxy contribution has been removed. The scatter is found to increase by another $\sim 0.4$ magnitude in the direction of attenuation and $\lesssim 0.2$ magnitude in the direction of specific star-formation (Appendix, Fig.~\ref{fig:uvj_clump_bkg_sub}). Hence the correlation between clumps and host galaxies still remains.

\subsection{Spatial distribution} \label{sec:spatial_distribution}

Due to the access we have to the rest-frame near-IR flux of our galaxy sample, we can accurately assess the distance of each clump from their host center. We do so by using the asymmetry centers, which also coincide with the location of bulges, of the galaxies in our sample (values borrowed from K24). Both are determined using the stellar light sensitive F444W image. To account for the different sizes of galaxies within our sample, we normalise the clump distance by the effective radius of the stellar disk in F444W, also from K24. The reason behind using the disk rather than the whole galaxy for the normalisation is to separate the bulge from this measurement. Moreover, if the central core increases in mass, it will decrease the net effective radius of the entire galaxy and hence artificially increase the normalised clump distance. 
\begin{figure*} 
    \centering
    \includegraphics[width=0.98\textwidth]{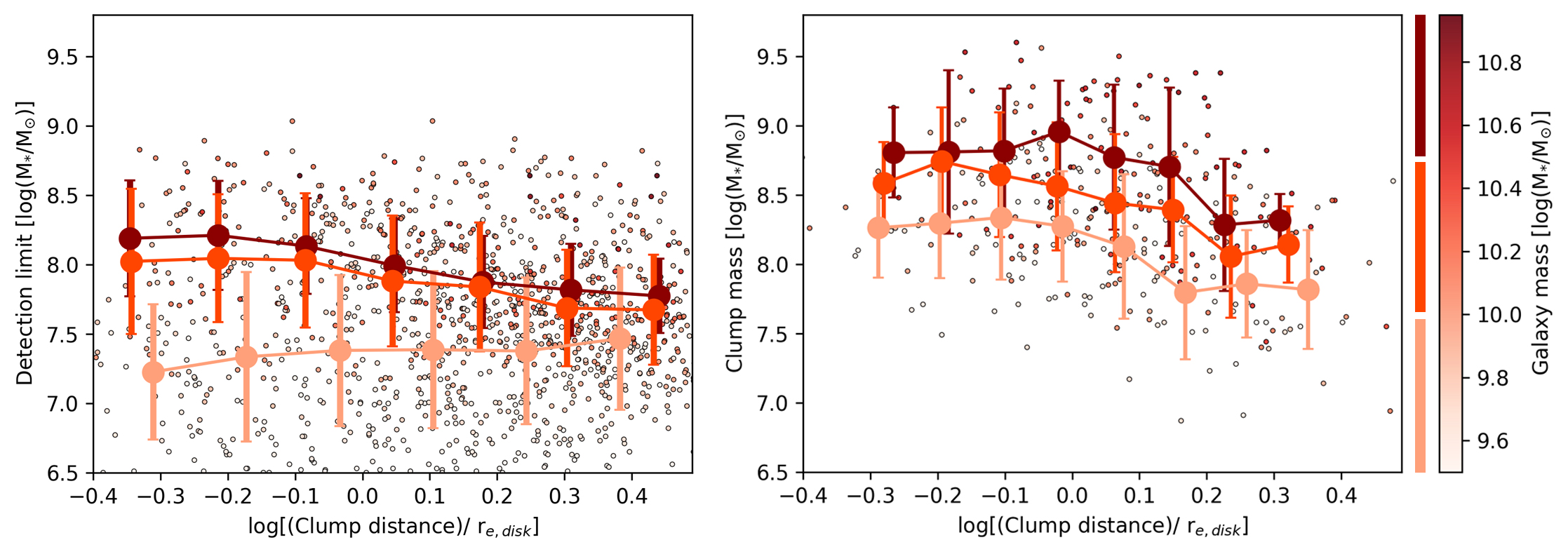}
    \caption{(Left) The effective mass of the detection limits for near-IR clumps as a function of the distance from the core of the host galaxy, normalised using the effective radius of the stellar disk from K24. (Right) The median of measured mass of the detected near-IR clumps within the galaxies as a function of the distance from the galactic center, found to be generally above the detection limits in the plot on the left. The continuous colorbar on the extreme right indicates the stellar mass of each clump (point) in the plot, while the three part bar alongside it shows the binned ranges used to measure the median values.}
    \label{fig:clump_mass_distance}
\end{figure*}

\begin{figure} 
    \centering
    \includegraphics[width=0.48\textwidth]{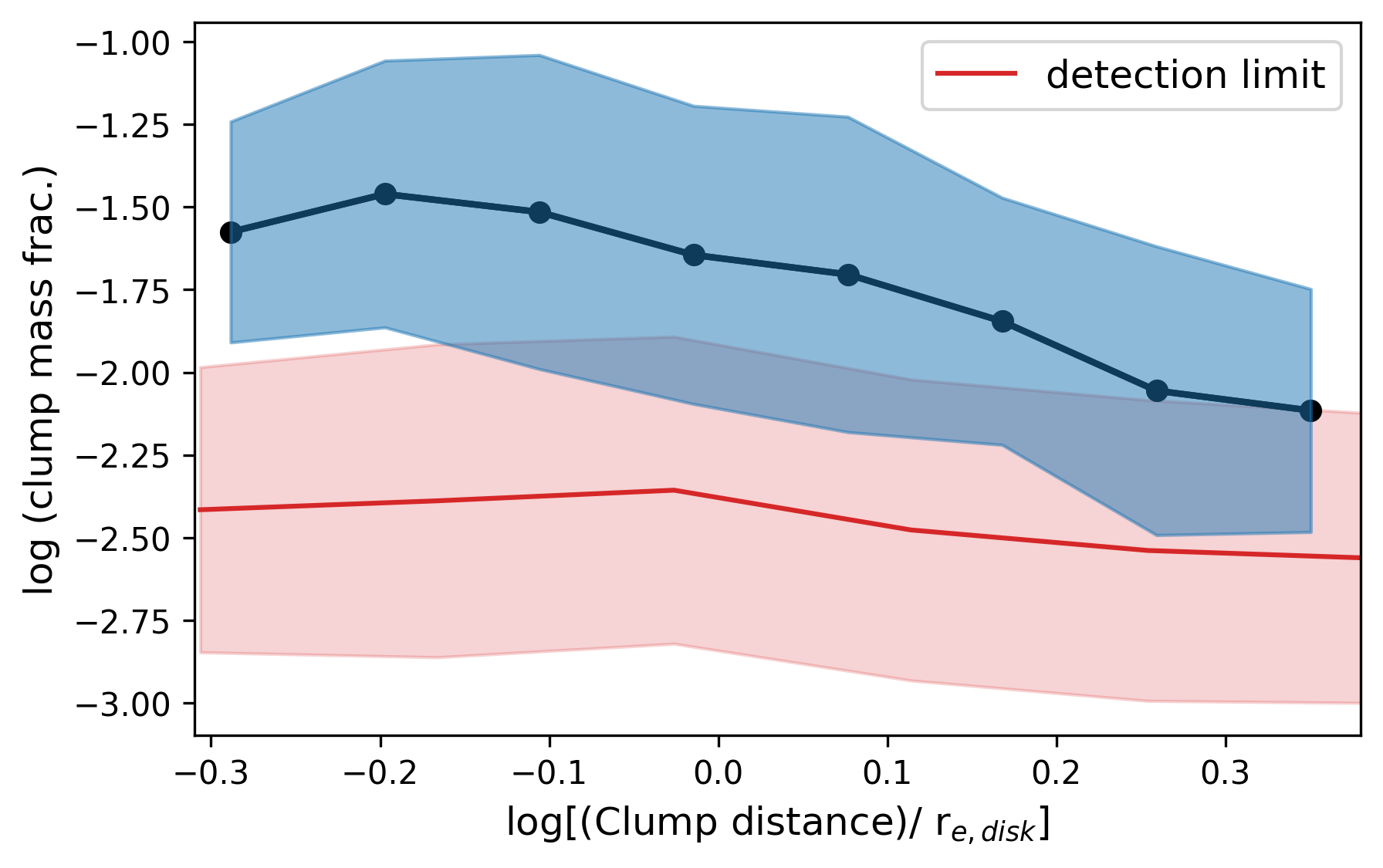}
    \caption{The ratio of stellar mass contained in clumps and that in the entire host galaxy, as a function of the normalised distance from the core of the host galaxy (as in Fig.~\ref{fig:clump_mass_distance}). The red line shows the estimation of the detection limits within the same parameter space.}
    \label{fig:clump_mass_frac_distance}
\end{figure}

\begin{figure*} 
    \centering
    \includegraphics[width=\textwidth]{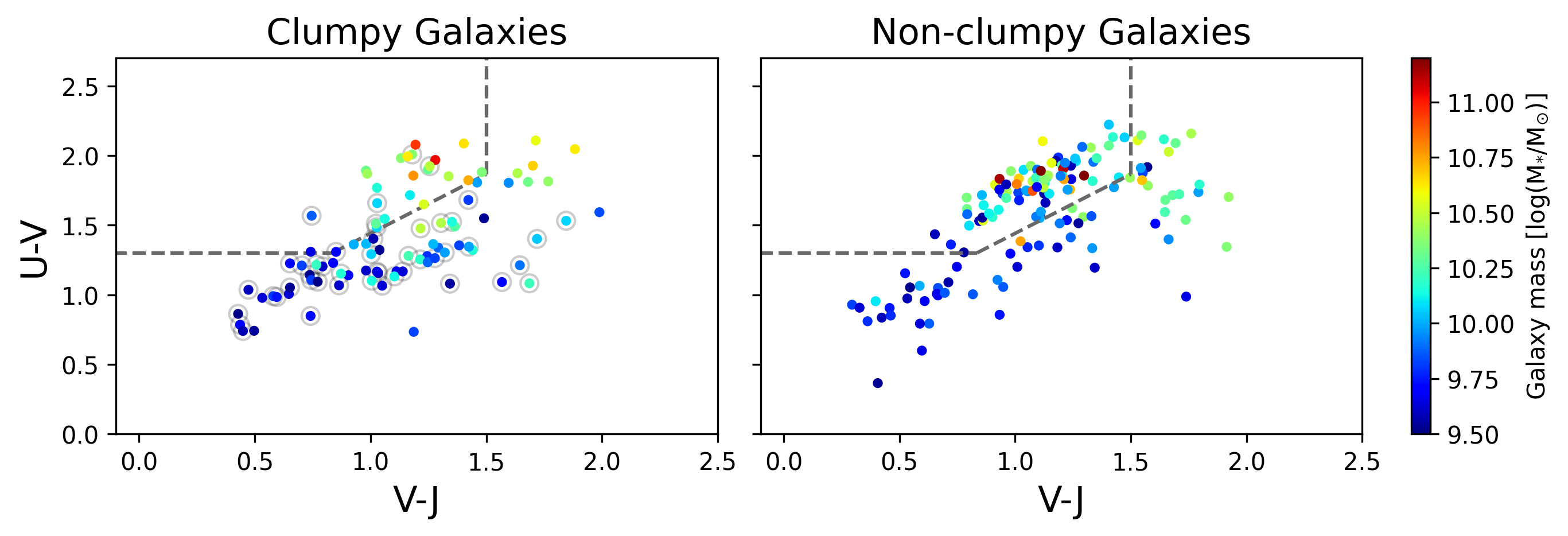}
    \caption{(Left) The UVJ plot for all galaxies within which clumps, either in UV or near-IR are detected. The color indicates the stellar mass of the host galaxy. The points with an additional black circle are the ones which have UV detected clumps. The dashed lines demarcate quiescent galaxies (region to the top-left). Moreover, shifting from the bottom-left to top-right involves an increase of dust attenuation. (Right) The same plot for the rest of the sample, which do not feature any clumps.}
    \label{fig:uvj_galaxy}
\end{figure*}

In Fig.~\ref{fig:clump_mass_distance} (right), we divide the clumps into three bins of host galaxy mass. This controls the rise of the detection limit as a function of galaxy mass indicated in Fig.~\ref{fig:clump_mass_plot}. Across the bins ($10^{9.5-10.0}\,\rm M_{\odot}$, $10^{10.0-10.5}\,\rm M_{\odot}$ and $10^{10.5-11.2}\,\rm M_{\odot}$), we see an increase of $\sim 10^{8.3-8.6}\,\rm M_{\odot}$, or $0.7\,$dex between the two extremes of our range of the distance from the core. It should be noted that we detect clumps only up to about 2.5 times the effective radius of the stellar disk, which reemphasise our use of the segmentation map from F444W to demarcate the region within the galaxy and not include neighbours. To ensure that the increase in clump mass with decreasing radial distance is not a systematic effect of the method used, with clumps closer to the centre being harder to detect, we create the same plot using the detection limits determined in Sec.~\ref{sec:mass_completeness} (Fig.~\ref{fig:clump_mass_distance}, left). We find that the maximum systematic increase in clump mass is of $\sim 0.3\,$dex for the highest galaxy mass bin, which is $\sim 0.4\,$dex lower than the increase in measured clump mass across the same distance range. The situation is even clearer in the lowest galaxy mass bin where we actually observe a reverse trend of decreasing limit, most likely a result of the scatter. 

While comparing the two plots in Fig.~\ref{fig:clump_mass_distance}, it is evident that the underlying density of datapoints in the plot for the detection limit decreases with decreasing radial distance. This is not observed in the plot on the right with the measured values. The reason behind it is simply that the actual length scale covered in the first half of the logarithmic x-axis is about 3 times smaller than the second half. Meanwhile, we follow a uniform distribution while artificially introducing clumps across the whole range. The plot reflects the gradient due to conversion from a linear to logarithmic scale. However, for the actual measurements, we detect the real clumps within the galaxy. The occurrence of these should approximately follow the morphology of the exponential disk, which has the same flux within and outside the effective radius by definition. Therefore, the density of clumps appears similar in both halves of the second plot. This clear distinction in clump density distribution between the two plots corroborates the idea of clumps forming from the disk itself, rather than having an external origin (further discussed in Sec.~\ref{sec:mass_discussion}).

Fig.~\ref{fig:clump_mass_frac_distance} provides a generalised representation of Fig.~\ref{fig:clump_mass_distance} by replacing the binning of the datapoints based on the host stellar mass with using the same value to normalise the clump mass. This clump mass fraction removes possible contributions of the mass evolution within bins, especially since the detection limit is found to correlate with the host mass (Fig.~\ref{fig:clump_mass_plot} and \ref{fig:clump_mass_distance}). Here too we observe a trend of increasing fractional mass in clumps with decreasing radial distance, with the detection limit clearly not responsible for it. Moreover, the detection limit mainly affects the proper sampling of the population of clumps further away from the center, flattening the net relation. Hence the trend could actually be steeper.

Therefore, we conclude that the property change in clumps as a function of distance from the galactic centre is robust against uncertainties and systematic effects from the methodology of this work. It could be suspected that the mass increase could be an effect of the subtraction method we incorporate. We check the corresponding distribution with the clump flux measured after the annulus subtraction method (Sec.~\ref{sec:clump_detection}). We find the result identical to Fig.~\ref{fig:clump_mass_frac_distance} but with a decrease in the normalisation by $\sim 0.1\,\rm dex$. We also recreate the exact same plot, but with the mass determined from flux measurements in the contrast images rather than the original images (provided in the Appendix, Fig.~\ref{fig:clump_mass_frac_distance_bkg_sub}). The result is not as identical, but it still shows the same increasing trend towards the core.

Given that the contrast between the clumps and the galaxy would decrease towards the core, the subtraction would artificially flatten this relation. As we still observe the same radial gradient (albeit with larger scatter), we conclude that the method of subtraction does not affect our results. Hence, we conclude that the the mass of clumps do effectively increase as a function of radial distance.

\subsection{The host galaxy population}

Fig.~\ref{fig:uvj_galaxy} shows the UVJ position for the population of galaxies within which we detect clumps, compared to those where we do not. We title these two sets as `clumpy' and `non-clumpy' galaxies respectively. The first clear distinction we observe is the fraction that can be classified as UVJ quiescent. This is found to be $23\,\%$ for clumpy galaxies and $59\,\%$ for those without clumps. Hence, clumpy galaxies are predominantly star-forming, with clumps being detected in either UV or near-IR. Furthermore, this fraction of UVJ quiescent galaxies goes down to $13\,\%$ if we only include sources showing UV clumps. We observe that there is a clear gradient with galaxy stellar mass, with less massive clumpy galaxies featuring low attenuation. A large fraction of them have UV clumps. In comparison, the non-clumpy population lacks this gradual trend, and $79\,\%$ of this group of galaxies more massive than $\sim 10^{10.1}\,\rm M_{\odot}$ appear in the quiescent regime, while the remaining few are in the highly dust-attenuated regime. This indicates a fundamental difference in their evolutionary trajectories being experienced by the two populations of galaxies.

\section{Discussion} \label{sec:discussion}

\subsection{The relevance of near-IR clumps} \label{sec:near_IR_detection_relevance}
For about two decades, the apparent clumpiness of high-z ($\gtrsim 1$) galaxies has been a major subject of interest, with the feature being observed in rest-frame UV. Although this established the increased star-formation in these structures, determination of their stellar mass was uncertain. Estimations could be made, but a stellar mass dependent detection was required to ensure that selection effects do not influence the results. In this work we exploit the rest-frame near-IR image from F356W, directly accessing the stellar-mass tracing flux, essentially circumventing such issues.

K24 already discussed the relevance of clumps detected across rest-frame optical and near-IR bands in regards to the overall stellar structural change of galaxies from being disk to bulge dominated. In this work, we find that $82\,\%$ of UV clumps are also detected as near-IR clumps based on their corresponding spatial overlaps, while it is only $28\,\%$ the other way round (Sec.~\ref{sec:clump_selection}). This suggests that the near-IR clumps, appearing in $40\,\%$ of all galaxies, forms a more ubiquitous group of structures compared to those detected in UV. Some of these near-IR clumps have low enough dust attenuation to also be detectable (or at least a part of them) in UV, indicated by the location of UV clumps as well as near-IR clumps with UV counterparts in the lower left part of the UVJ diagram (Fig.~\ref{fig:uvj_clumps}). Whereas others can be classified as highly dust-obscured or quiescent based on the same UVJ colors. 

Furthermore, we make a visual inspection of each contrast map to understand the regions not selected as clumps. We find that in a large fraction of cases, one can easily discern more extended but less concentrated stellar structures, with the detected clumps forming the most concentrated regions. Some of these structures also resemble spiral-arms, but we are not in the position to quantify these features and therefore hold off any conclusions in this regard. Nevertheless, it is important to understand that the detected clumps in many instances are not isolated structures. This is a key perspective added by the near-IR detection.

\subsection{Mass of the clumps and galaxy subtraction} \label{sec:mass_discussion}

The clump mass range of $10^{7.5-9.5}\,\rm M_{\odot}$ with an average value of $\sim 10^{8.5}\,\rm M_{\odot}$ (Fig.~\ref{fig:clump_mass_plot}, top) measured in this study is found to be in general agreement with previous works \citep{wuyts12, guo12, guo18, huertas-company20}. The corresponding fraction of galaxy stellar mass contained in clumps of $1-3\%$ is sufficient to be driving gas towards the core through tidal torques \citep{bournaud14}. Hence, irrespective of the fate of the clumps and their exact structures, the level of mass concentration we measure establishes a consequential effect on galaxy evolution through enhancement of centralised gas concentration. Nevertheless, there may be clumps of lower masses which are simply below our detection threshold. This is also indirectly related to the scales we are sensitive to, the lower limit determined by the PSF size $\sim 0.06^{\prime\prime}$ and the upper limit dictated by the smallest scales we subtract $\sim 0.14^{\prime\prime}$. In physical scales, these will be between $0.5-1.2\,$kpc for our redshift window. 

\begin{figure} 
    \centering
    \includegraphics[width=0.48\textwidth]{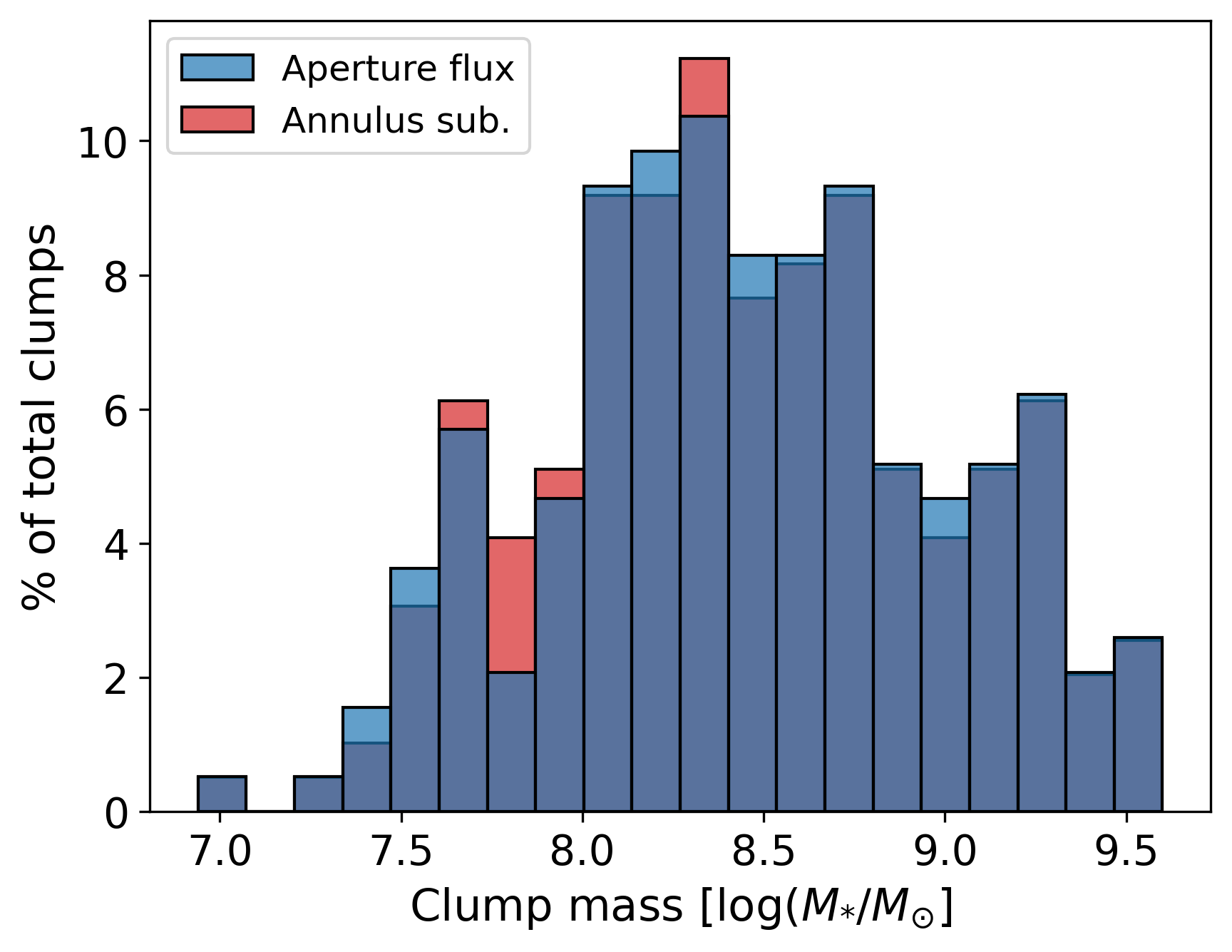}
    \caption{Histogram showing the clump mass distribution within our sample using the two methods of flux measurement, using the flux within the region of the clump, and the same after subtracting the annulus background flux.}
    \label{fig:clump_mass_histogram}
\end{figure}

As discussed in Sec.~\ref{sec:clump_detection}, we remeasure the clump flux in each filter by subtracting a background value, estimated using an annulus at the radial distance of the respective clump. This process leads to an over-subtraction in some cases (sec.~\ref{sec:mass_completeness}), leading to an inability to estimate detection limits. Nevertheless, we find that the resulting clump mass distribution remains almost the same (Fig.~\ref{fig:clump_mass_histogram}) with an incremental decrease of $\sim 29\%$ ($0.1-0.2\,\rm dex$) in the median value of the sample. The upper limit of the clump mass distribution ($\sim 10^{9.5}\,\rm M_{\odot}$) remains the same however. Furthermore, we find no discernible difference in other results. Therefore, we support the use of the aperture method that we have primarily discussed for which we have robust detection limits. 

Background subtraction has been implemented with varying recipes in previous works \citep{forster11, wuyts12, guo12, guo18}, which have targeted rest-frame UV. Some also explored the effects different methods may have and found that the flux values may change up to a factor of $\sim 3$ using different methods with a corresponding UV flux reduction of $\sim 15-30\,\%$ \citep{forster11, guo12, guo18} as a result of the subtraction. In this work, we find a similarly minimal decrease in the near-IR flux reflected in the stellar mass estimates. This difference hence does not translate to any detectable offset in the clump mass estimates between our results and previous works. However, it is prudent to point out that over-subtraction can especially impact the rest-frame near-IR bands with lower contrast between clumps and the background. This makes background estimation more challenging. 

\subsection{Hierarchical nature of clumps}

The cSMF provides key insights into the formation and evolution of clumps in galaxies. Theoretical models of turbulence-driven clump formation predict a slope of -2 \citep{dessauges18, elmegreen18}. A similar slope also suggests a hierarchical nature, comparable to that observed in star-forming regions of local galaxies \citep{elmegreen06,veltchev13,chandar14,adamo17,zhou24}. However, the slope is expected to become shallower as clump populations evolve. Additionally, resolution can flatten the cSMF due to blending effects \citep{huertas-company20}. However, these effects are insufficient to fully alter the stellar mass distribution \citep{tamburello17}, as long as the cSMF is estimated for clumps above the detection limit of the study.

Our cSMF estimation of $-1.50 \pm 0.14$ agrees with values observed for lensed galaxies in \cite{dessauges17} within their error bars. However, the shallower profile likely reflects the resolution limits of our study, as well as possible high-mass bias introduced by our detection process (Sec.~\ref{sec:mass_completeness}). Fig.~\ref{fig:clump_lum_fucntion} shows that intermediate- to low-mass clumps are less likely to be included, compared to studies with less stringent selection criteria. Additionally, near-IR detection allows for more dust-obscured, and possibly more massive clumps (indicated in Fig.~\ref{fig:clump_mass_plot} and \ref{fig:uvj_clumps}). Effects of observational limitations were discussed in \cite{huertas-company20}, where they used galaxies from VELA simulations and replicated the systematic conditions of CANDELS images. For their full sample (VELA \textit{Candelized}), they find a slope similar to that in our study ($-1.55 \pm 0.34$). Improvements in resolution are likely to steepen the slope of the cSMF as shown in \cite{kalita24c}. They used JWST/NIRCam F150W rest-frame optical images (with three times higher resolution than F444W), and a robustness test based on the fitting of composite morphological models to get a slope of $-1.85 \pm 0.14$. 

Nevertheless, our measured slope suggests a hierarchical nature for clumps. Furthermore, \cite{dessauges17} proposed that the higher clump mass with increasing galaxy mass could result from clump mergers, giving rise to a hierarchical population. The same trend observed in our study (Fig.~\ref{fig:clump_mass_plot}) further substantiates this conclusion. This could also explain the radial variation of clump mass (sec.~\ref{sec:spatial_distribution}), with clumps at smaller orbits close to the center of the galaxy likely to undergo more mergers.

\subsection{Similarity with the host properties} \label{sec:sim_galaxy_clumps}

Directing the focus on to the host galaxies, we observe that they predominantly occupy the UVJ star-forming regime. Clumpy galaxies (in rest-frame UV) being star-forming and preferentially low-mass has been widely suggested \citep[e.g.,][]{guo15}. This is clear from Fig.~\ref{fig:uvj_galaxy} when we limit ourselves to those featuring UV clumps. However, if we include the entire sample of clumpy galaxies, including those featuring clumps in near-IR, we observe that a fraction of these also populate the high-attenuation or quiescent regimes in the UVJ diagram. In contrast, the galaxies without any clumps are dominantly quiescent. 

The distribution of the host galaxies in the UVJ plot resembles the same for the near-IR clumps. In both cases, we also observe a similar shift towards the top right of the plot with increasing host stellar mass, which indicates correlated increase in levels of attenuation or quiescence. This similarity is further illustrated in Fig.~\ref{fig:uvj_clumps_galaxy_diff}, and as discussed in Sec.~\ref{sec:clump_classification}, the scatter suggests similar levels of sSFR in clumps and galaxies across our sample. However, it is the dust attenuation that may vary to a slightly higher degree based on the increased scatter along the bottom-left to top-right direction.

\subsection{Possible evidence of migration}

Multiple works have suggested a radial change in clump properties to be an indication of migration into the core \citep{guo12, shibuya16, guo18, soto17, huertas-company20}. Our results presented in Sec.~\ref{sec:spatial_distribution} contribute to this discussion. We observe an increase of clump stellar mass with decreasing radial distance, even after taking into account systematic detection effects (Fig.~\ref{fig:clump_mass_distance} and \ref{fig:clump_mass_frac_distance}). We can hence test the feasibility of the migration scenario using the radial increment in stellar mass. Given that we do not extract values of star-formation rates from either the galaxies or individual clumps, we rely on the star-forming main-sequence relation from \cite{leslie20}. We assume that the galaxies predominantly follow it at least during their star-forming phase. If the mass increase of the clumps are also governed by the same relation, we find that we get the same increment (with star-formation rate $\sim 0.1$ dex above the relation but within the scatter) seen at the two extremes of both Figs.~\ref{fig:clump_mass_plot} and \ref{fig:clump_mass_distance}. This takes into account the expected migration timescales $\sim 0.4-1.0\,$Gyr \citep{dekel22}. We can hence suggest a scenario where the clumps follow the galaxy in their star-formation properties while also migrating inward. This leads to the more massive clumps ($\gtrsim 10^{8.5}\,\rm M_{\odot}$) being closer to the center, with the less massive ones not surviving disruption due to stellar feedback. 

However, the clumps may not follow the star-forming main sequence to produce an exponential rise in stellar mass. \cite{dekel22} concludes that the star-formation rates rather remain constant as it is dictated by the accretion rate of the clump. However, this difference would likely be within the scatter observed in Fig.~\ref{fig:uvj_clumps_galaxy_diff}. Neither of the two scenarios can be rejected based on our SED-based SFR estimations (Appendix, Fig.~\ref{fig:sed_sfr}). Moreover, the constant SFR narrative would also have to contend with the variation we observe in the UVJ classification of near-IR clumps, with the mass increase of the clumps resulting in the various positions in the UVJ plot. Although, if one only considers the UV clumps we observe in our sample, the UVJ positions are much less scattered. More accurate SFR indicators would be required to resolve this question of SFR variations in migrating clumps.  

On the other hand, the mass gradient may simply be an effect of higher concentration of gas near the core \citep[e.g.,][]{elbaz18, puglisi21, gomez-guijarro21} and the clumps actually would not survive long enough to migrate across the galactic disk. This would drive higher levels of star-formation dictated by the well established Schmidt-Kennicutt relation \citep{kennicutt98}. However, that would imply the stellar mass gradient observed in Fig.\ref{fig:clump_mass_distance} is simply due to the inner and outer regions forming stars at different rates. The difference would hence be created over periods much shorter than the migration timescales of clumps. Assuming that the initial mass of the clumps are generally similar, an approximate difference in specific SFR of $\gtrsim 0.4\,$ dex would be required if one implements an extrapolation of the star-forming main-sequence relation from \cite{leslie20}. This difference is well within the scatter of both the SED-based SFR and specific SFR distribution of clumps as a function of radial distance from the centre (plots shown in Appendix, Fig.~\ref{fig:sed_sfr}).

We further investigate the stellar ages and dust attenuation across the surface of the galaxies to understand better the reasons of this apparent increase in clump mass with decreasing distance from the centre. As mentioned earlier in Sec.~\ref{sec:fast_sed}, the exact values of these parameter may not be robust enough to include in the conclusions of this work. Nevertheless, we mention them here primarily for a discussion and the corresponding plots are provided in the Appendix (Fig.~\ref{fig:sed_age_av}). We find that the stellar ages marginally decrease towards the core of the galaxy, along with an increase in the attenuation. However, it is also possible that we are being influenced by the dust-age degeneracy which can artificially reduce the stellar age through an increase in attenuation. However, even if one assumes that the stellar age is being underestimated near the core, we would expect a further increase in the stellar mass of the clumps in the region. This would further enhance the gradient in Fig.~\ref{fig:clump_mass_distance}, albeit marginally. 

Therefore, we can confirm a radial gradient in the clump properties based on the SED-based robust stellar mass along with the less reliable stellar age and dust attenuation measurements. Nevertheless, detailed sub-mm resolved observations measuring the IR-based SFR or resolved spectral line SFR measurements would be critical in understanding the key driver of the observed radial trend in clump mass. Such observations would also allow for a quantification of the mass build-up in galaxies and what role clumps play, which is intrinsically connected to the question of migration.

\subsection{The nature of detected clumps}

Besides the advantages of detecting clumps in the stellar mass tracing near-IR light that have been discussed in this work, it is important to highlight the caveats. Rest-frame UV studies of lensed galaxies suggest that the the intrinsic sizes of clumps are a $\sim 100\,\rm pc$ with masses on average $\sim 5\times 10^{7}\,\rm M_{\odot}$ \citep{dessauges17, soto17}. Whereas, direct observations of non-lensed galaxies have found sizes $\sim 1\,\rm kpc$ and stellar masses higher by about an order of magnitude \citep{elmegreen07, guo15, guo18, huertas-company20}. Comparisons to local galaxies do not resolve this discrepancy as their clumps are found to be even lower in mass ($\sim 10^{5}\,\rm M_{\odot}$) and a large range of sizes from $\sim 10\,\rm pc$ to $\sim 1\,\rm kpc$ \citep{kennicutt03, bastian05}.

Unsurprisingly, our results from the near-IR study of clumps returns similar results to those predicting massive clumps of $\sim 1\,\rm kpc$ sizes since the resolution of our images are similar to the UV band images that have been previously used. However, as discussed in Sec.~\ref{sec:near_IR_detection_relevance}, using the near-IR comes with the added advantage of creating a stellar mass based detection. Therefore also detecting clumps which may be highly obscured or have older stellar populations, leading to suppressed UV flux. However, this can also lead to our venturing into gravitationally bound over-dense regions including spiral arms or tidal features. Complicating matters further, simulations suggest that clumps may more likely form at location of stellar substructures, especially in cases of lower gas fractions \citep{frensch21}.

The similarity between near-IR clumps and their host galaxies  (Sec.~\ref{sec:sim_galaxy_clumps}) further adds to the question about the nature of these kpc-scale `clumps'. If UV clumps are primarily star-forming regions, the near-IR clumps likely indicate associated stellar structures, potentially dense substructures forming within the disks. This brings into question the contribution of classical clumps, formed from disk instabilities, to the stellar substructures of the galaxy disk \citep{kalita24c,kalita25}. Further investigation is needed to fully understand the distinction (and overlaps) between clumps and substructures.

Therefore, we have to reassess the relevance of our definition of clumps. This work has mainly referred to any clumpy structure appearing to have a higher surface density than its surroundings as clumps, similar to the recipe used in previous UV-based studies. However, based on lensed galaxy studies we can expect that a sizeable fraction of these are rather clump `clusters' made up of structures smaller by an about order of magnitude. Our results in near-IR suggests that irrespective of the smallest building block of these structures, the net mass concentrated within their $\sim 1\,\rm kpc$ size is $10^{7.5-9.5}\,\rm M_{\odot}$. Hence, these structures are theoretically massive enough to drive structural evolution of the host galaxy by driving gas inward, corroborated by observational evidence in K24. The clumps that were previously detected in rest-frame UV make up a part of this group. However, there are additional structures which also do exist but happen to feature suppressed UV flux.

\subsection{Clumpy vs non-clumpy galaxies}

Based on the SED fitting results of the galaxies, with and without clumps, it is clear that clumpy galaxies are preferentially star-forming. In comparison, the majority of non-clumpy galaxies are UVJ quiescent. Focusing on the clear stellar mass gradient we observe for the clumpy galaxies in Fig.~\ref{fig:uvj_galaxy}, one could suggest this trend is indicative of an evolving star-forming galaxy, with rising dust content with increasing mass \citep{fang18}. Along with the host, the co-evolving clumps within it also increase in mass and dust attenuation. How this gradual change in UVJ characteristics is related to the evolution in clumpiness as a function of the galaxy morphology (K24) will have to be understood in future studies. 

The non-clumpy population, in contrast seem to be categorised as quiescent much more predominantly and is likely different physical conditions compared to the clumpy galaxies. K24 links the low clumpiness with bulge dominance and hence our results could indicate a bulge-driven morphological quenching. We study the difference between the two population in a follow-up work (Kalita et al. in prep) 

\section{Summary and conclusion} \label{sec:conclusion}

In this work we exploit the wide spectral coverage ($0.5-4.6\,\mu$m) of a stellar mass complete sample of galaxies at $1 < z < 2$ to study clumps within them. We combine CEERS public data from both HST/ACS and JWST/NIRCam to make this possible. Throughout the wavelength range, we have sufficient spatial resolution to properly resolve the expected $\sim 1\,\rm kpc$ clump sizes. We use a wavelet decomposition method to create contrast maps in both rest-frame UV (F606W) as well as near-IR (F356W) that allows for an automated clump detection strictly within the stellar extent of the respective host galaxies. 
\begin{itemize} [leftmargin=*, noitemsep, topsep=0pt]
    \item We find the total fraction of galaxies showing near-IR clumps is $40\%$. Amongst the UV clumps, $85\%$ have a spatial overlap with a near-IR clump. Meanwhile only $28\%$ of near-IR clumps have a UV counterpart. We suggest that a stellar-mass tracing near-IR detection of clumps provides a larger and more complete sample of clumps in galaxies compared to UV. The reason is primarily due to the effects of dust attenuation and the star-formation influencing the UV flux. 
    \item Using the rest-frame near-IR for detection allows for a determination of the detection limit in terms of stellar mass across our sample. We find that a similar estimation is not possible with an UV detection due to a widely varying SED influenced by dust attenuation and stellar age. 
    \item The near-IR clumps closely follow the UVJ characteristics of the host galaxy, thereby indicating a similarity in sSFR and dust attenuation rates. Furthermore, the hosts (and hence their clumps) primarily lie in the star-forming regime. A fraction of them ($23\%$) are located in the UVJ quiescent region however, a characteristic also shown by near-IR detected clumps ($16\%$ are found to be UVJ quiescent). 
    \item The stellar mass of clumps within our sample is found to be in the range of $10^{7.5-9.5}\,\rm M_{\odot}$ with an average value $\sim 10^{8.5}\,\rm M_{\odot}$. We find the detection limit in F356W filter to correspond to a mass of $\sim 10^{8.1}\,\rm M_{\odot}$. Furthermore, above this detection limit, we find the cSMF to follow a power law with $\alpha = -1.50 \pm 0.14$. This value indicates the hierarchical nature of clumps likely a result of growth through clump mergers.
    \item We also find a clear increase in the clump mass with decreasing distance from the near-IR based location of the galaxy center. This radial trend is also reflected in the fraction of galaxy stellar mass within clumps. We find both these correlations robust against selection effects. The less reliable stellar age and dust attenuation derived from SED fitting also corroborate this conclusion. This result could suggest inward migration of the clumps, although accurate star-formation tracers would be required to confirm it.  
    \item Throughout, we have quoted results derived by measuring clump fluxes directly on the original images. However, we find almost perfect agreement of the results had the fluxes been measured after an radial annulus based background subtraction. There is only a reduction of the median
    of the clump mass distribution in our sample by 29\%, although
    without changing the upper limit of the distribution. We conclude that not subtracting the background provides more reliable results given that we are not affected by varying levels of over-subtraction along with the possibility of determining reliable detection limits.
\end{itemize}

\section*{Acknowledgements}

We thank the referee for the crucial comments and suggestions. B.S.K. and J.D.S. are supported by the World Premier International Research Center Initiative (WPI), MEXT, Japan. J.D.S. is supported by JSPS KAKENHI (JP22H01262). This work was also supported by JSPS Core-to-Core Program (grant number: JPJSCCA20210003). W.M. acknowledges the funding of the French Agence Nationale de la Recherche for the project iMAGE (grant ANR-22-CE31-0007). L.C.H. is supported by the National Science Foundation of China (11721303, 11991052, 12011540375, 12233001), the National Key R\&D Program of China (2022YFF0503401), and the China Manned Space Project (CMS-CSST-2021-A04, CMS-CSST-2021-A06)

\section*{Data Availability}

The data underlying this article are publicly available in the online archives of the CEERS collaboration (https://ceers.github.io/releases.htm)



\bibliographystyle{mnras}
\bibliography{main} 




\appendix

\section*{Determining wavelet scales for subtraction}

We decide on the smallest scale (and everything greater) to be subtracted by making comparisons to the previous method used in K24. A rough estimation by visual inspection of subtraction residuals of clumpy galaxies, along with only subtracting upto scales that are larger than the image PSF, indicates this scale to be somewhere between $0.3^{\prime\prime}$ and $0.1^{\prime\prime}$. Among our seven wavelet scales, this leaves us three possibilities: $0.27^{\prime\prime}$, $0.14^{\prime\prime}$ and $0.08^{\prime\prime}$. These three options are shown as Sub. A, B and C respectively in Appendix Fig.~\ref{fig:gal_sub_comp} under the column of Method [B]. Method [A] refers to the previous method of smoothing using a Gaussian kernel of $\sigma = 3\,\rm pixels$ as done in K24.

We find that only removing scales upto $0.27^{\prime\prime}$ (Sub.~A), leads to higher negative peaks near emission regions in the residual image, compared to method [A]. This can also be deduced from the difference image, where these appear as positive flux. We hence conclude that this is not a sufficient background removal. On the other hand, subtracting scales equal to and greater than $0.08^{\prime\prime}$ leads to much less negative residuals. However, we find that the clumps hence detected do not include a large fraction of the flux compared to method [A]. This is found to be significantly higher than the flux measurement uncertainties\footnote{the uncertainties for these test cases are borrowed from the values measured in K24 for the specific sources} for each clump. Moreover, through visual inspection we also observe that this translates to clump getting more difficult to detect. In other words, this aggressive subtraction severely reduces our ability to detect clumps in our sample galaxies.

We therefore settle on subtraction~B of scales upto $0.14^{\prime\prime}$. In this case, the negative residuals are also found to be lower than the previous method, thereby improving the background subtraction (by $\sim 30-40\%$). Meanwhile the clump fluxes, are only lower by an amount much less than the flux uncertainties. Hence detection of clumps remain at a similar level.  

\begin{figure*} 
    \centering
    \includegraphics[width=\textwidth]{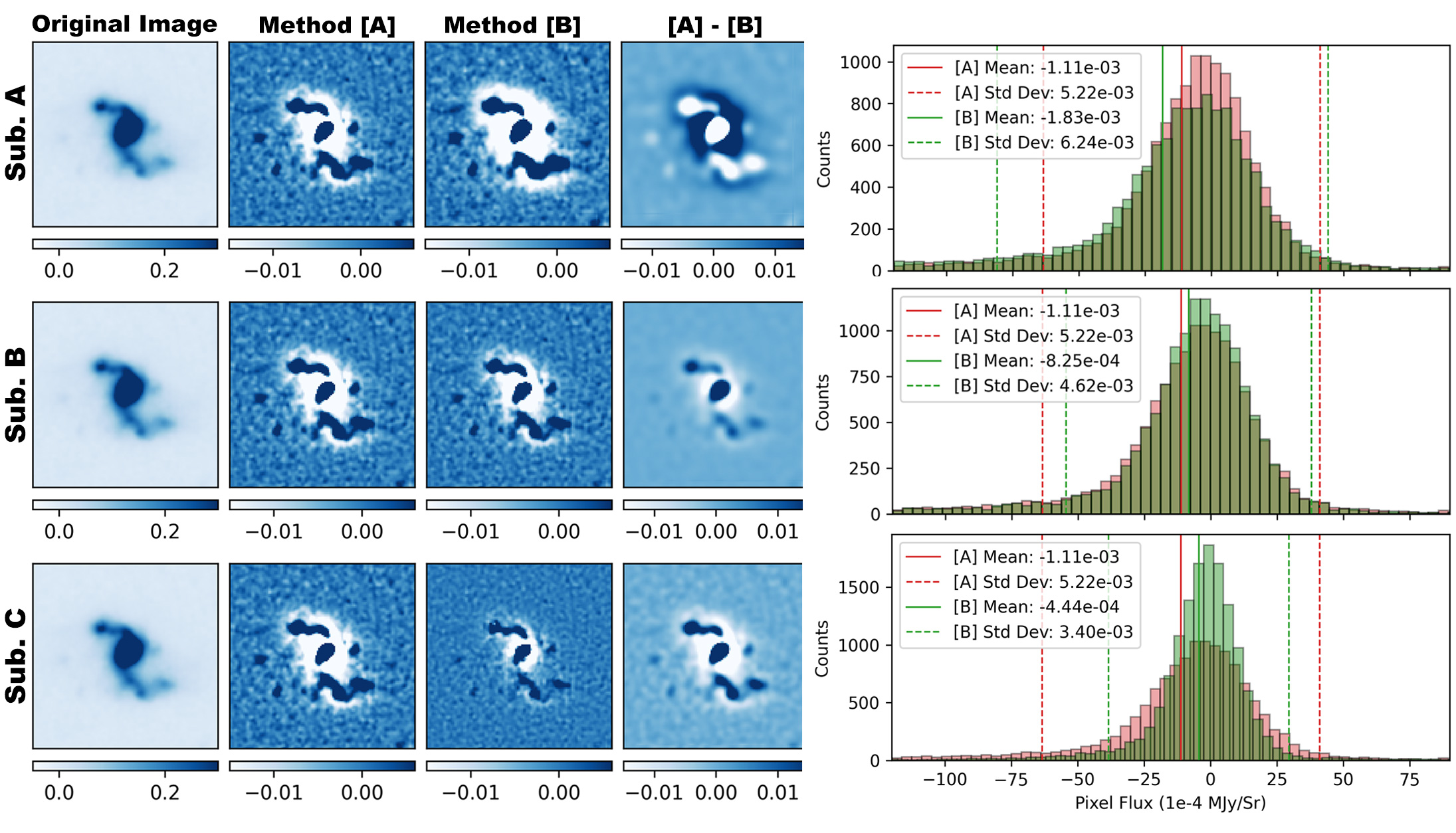}
    \caption{A comparison between two methods of galaxy subtraction}
    \label{fig:gal_sub_comp}
\end{figure*}

\section*{Contrast image property limits}
\begin{figure*} 
    \centering
    \includegraphics[width=0.9\textwidth]{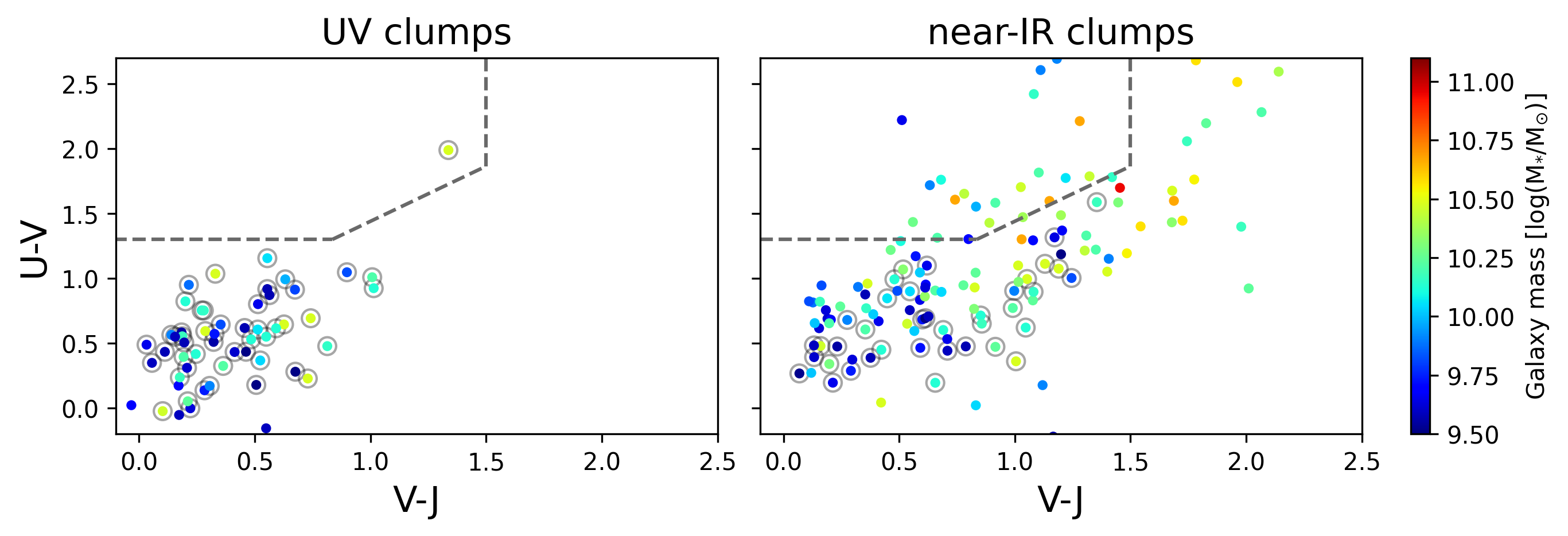}
    \includegraphics[width=0.9\textwidth]{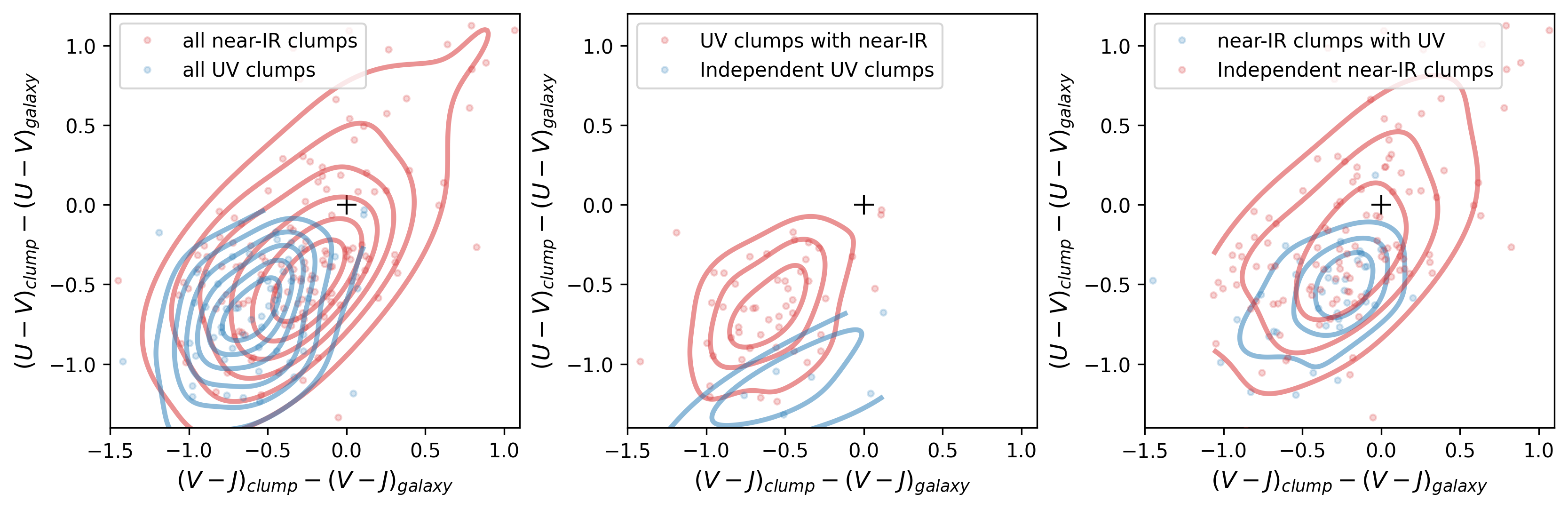}
    \caption{The UVJ plot (top) for clumps as well as their difference in both axes to the respective host galaxies (bottom). These are generally similar to the version of the results in the main text, albeit with higher scatter by up to $\sim 0.4$ dex.}
    \label{fig:uvj_clump_bkg_sub}
\end{figure*}

\begin{figure*} 
    \centering
    \includegraphics[width=0.48\textwidth]{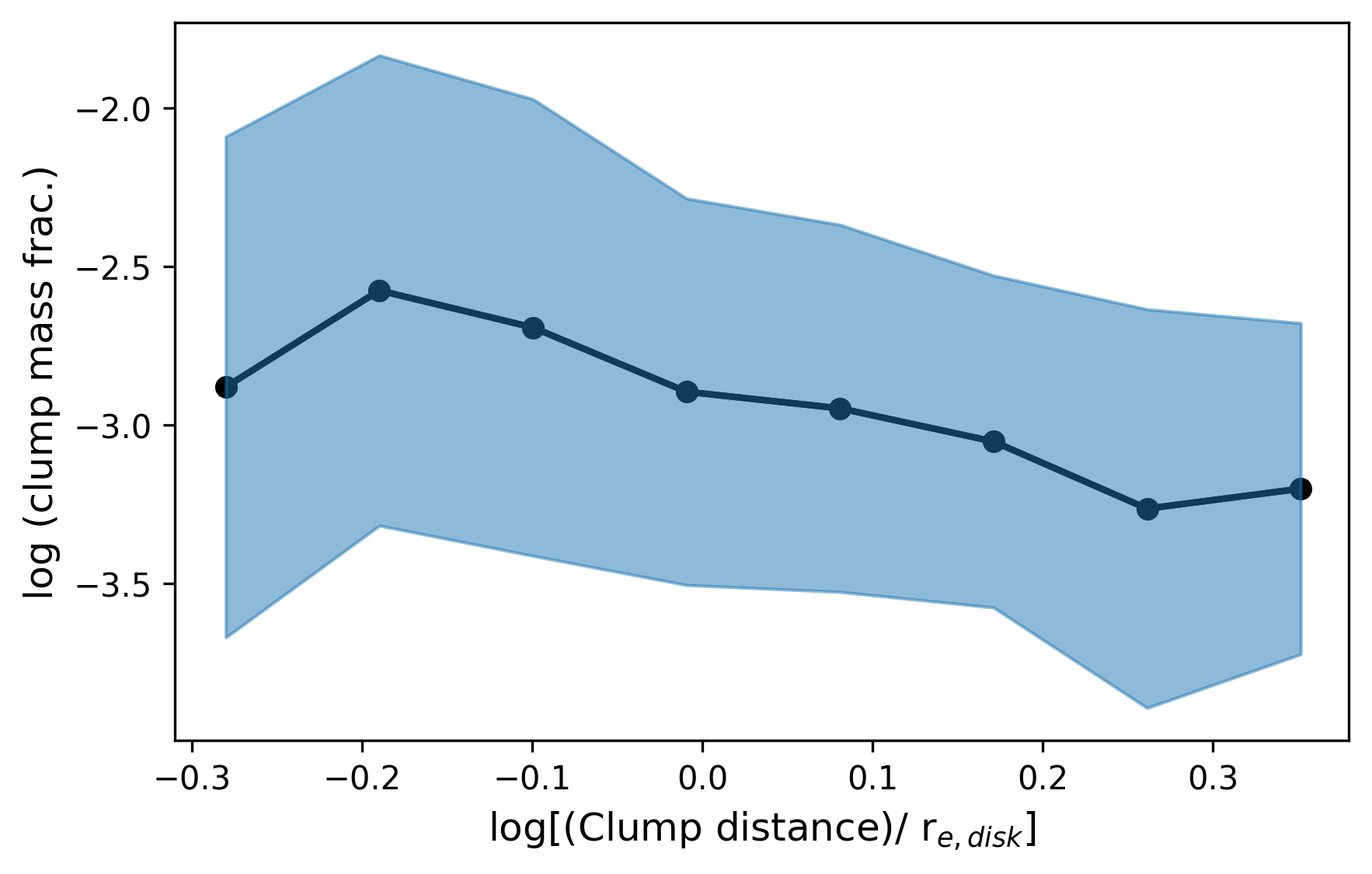}
    \caption{The version of Fig.~\ref{fig:clump_mass_frac_distance} with the clump mass estimations derived from the contrast images in each filter rather than the original images. The relation appears flatter than that in Fig.~\ref{fig:clump_mass_frac_distance}, likely due to reasons discussed in Sec.~\ref{sec:mass_completeness} in addition to possible over-subtraction near the core which is also indicated in Fig.~\ref{fig:wavelet_decomp}.}
    \label{fig:clump_mass_frac_distance_bkg_sub}
\end{figure*}

Although we support the use of the original images for clump flux measurements (Sec.~\ref{sec:mass_completeness}), here we provide the version of results of this work if the contrast images were used (Appendix Fig.~\ref{fig:uvj_clump_bkg_sub}, \ref{fig:clump_mass_frac_distance_bkg_sub}). These should be regarded as limits of the measurements and are shown as evidence of the robustness of the results presented in the main text.

\section*{Additional SED based properties}

\begin{figure*} 
    \centering
    \includegraphics[width=0.48\textwidth]{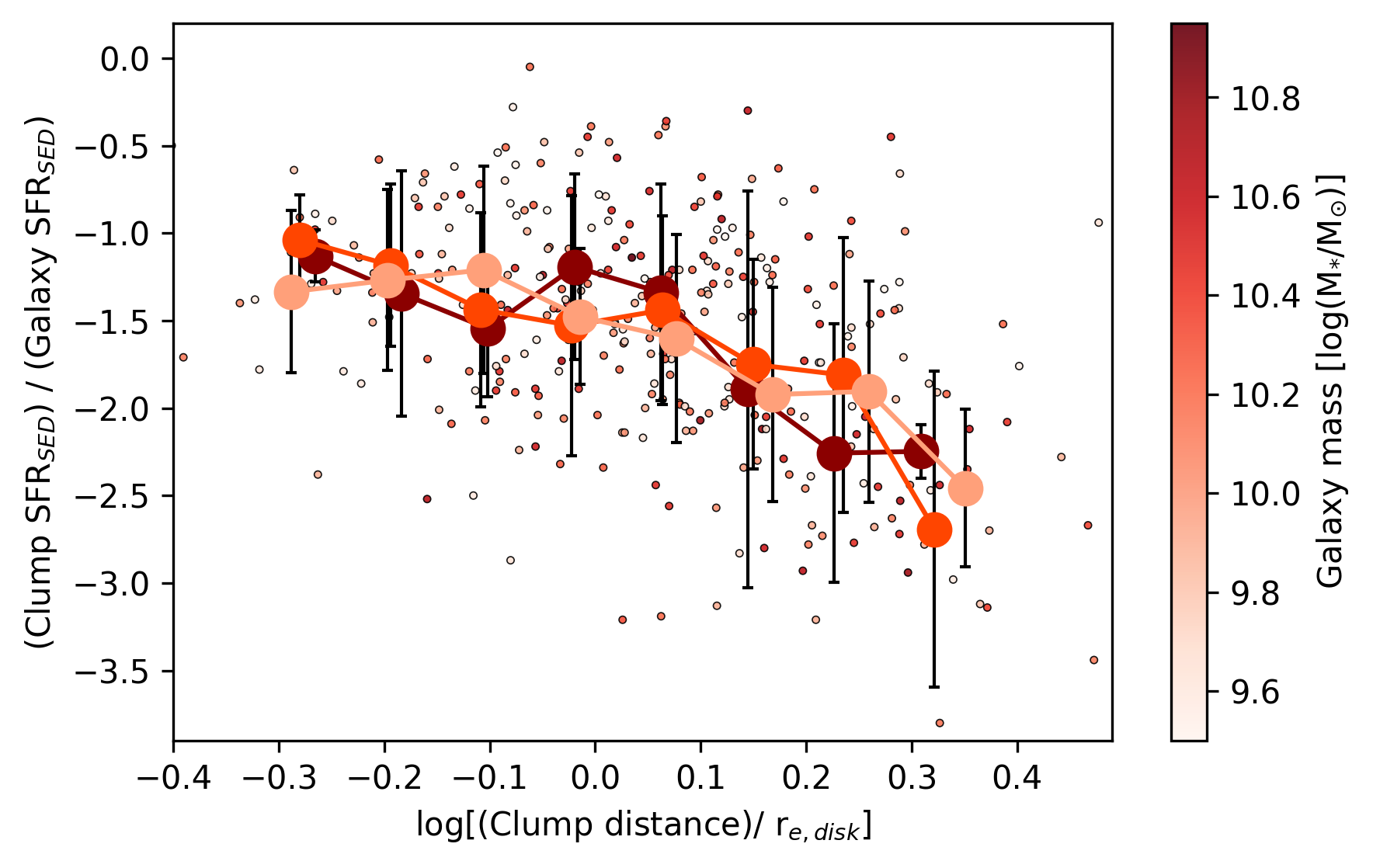}
    \includegraphics[width=0.48\textwidth]{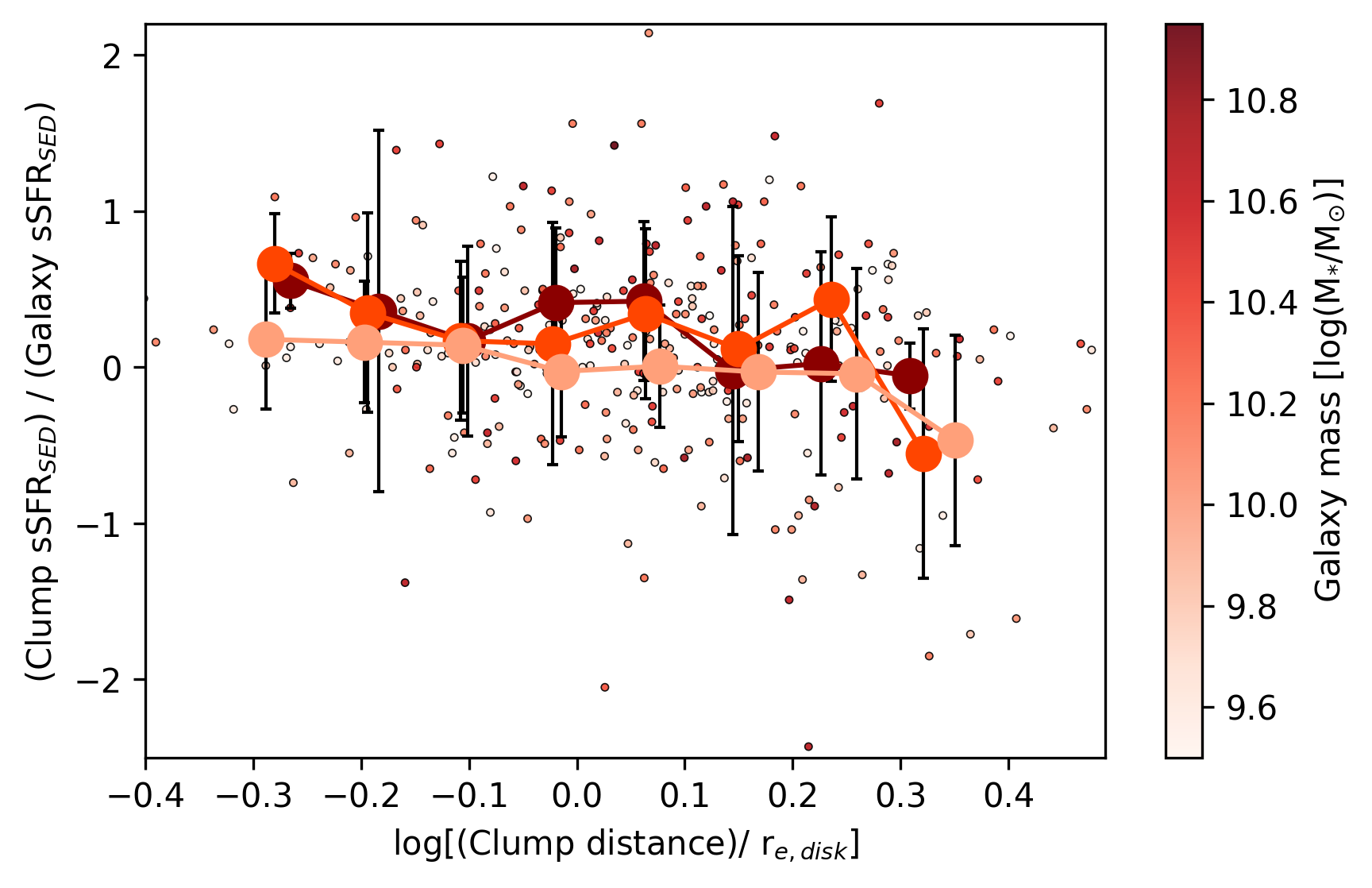}
    \caption{The SED-derived median SFR and specific SFR (SFR/stellar mass) in log scales, on the left and right respectively, as a function of distance from the galactic core across our sample.}
    \label{fig:sed_sfr}
\end{figure*}

\begin{figure*} 
    \centering
    \includegraphics[width=0.48\textwidth]{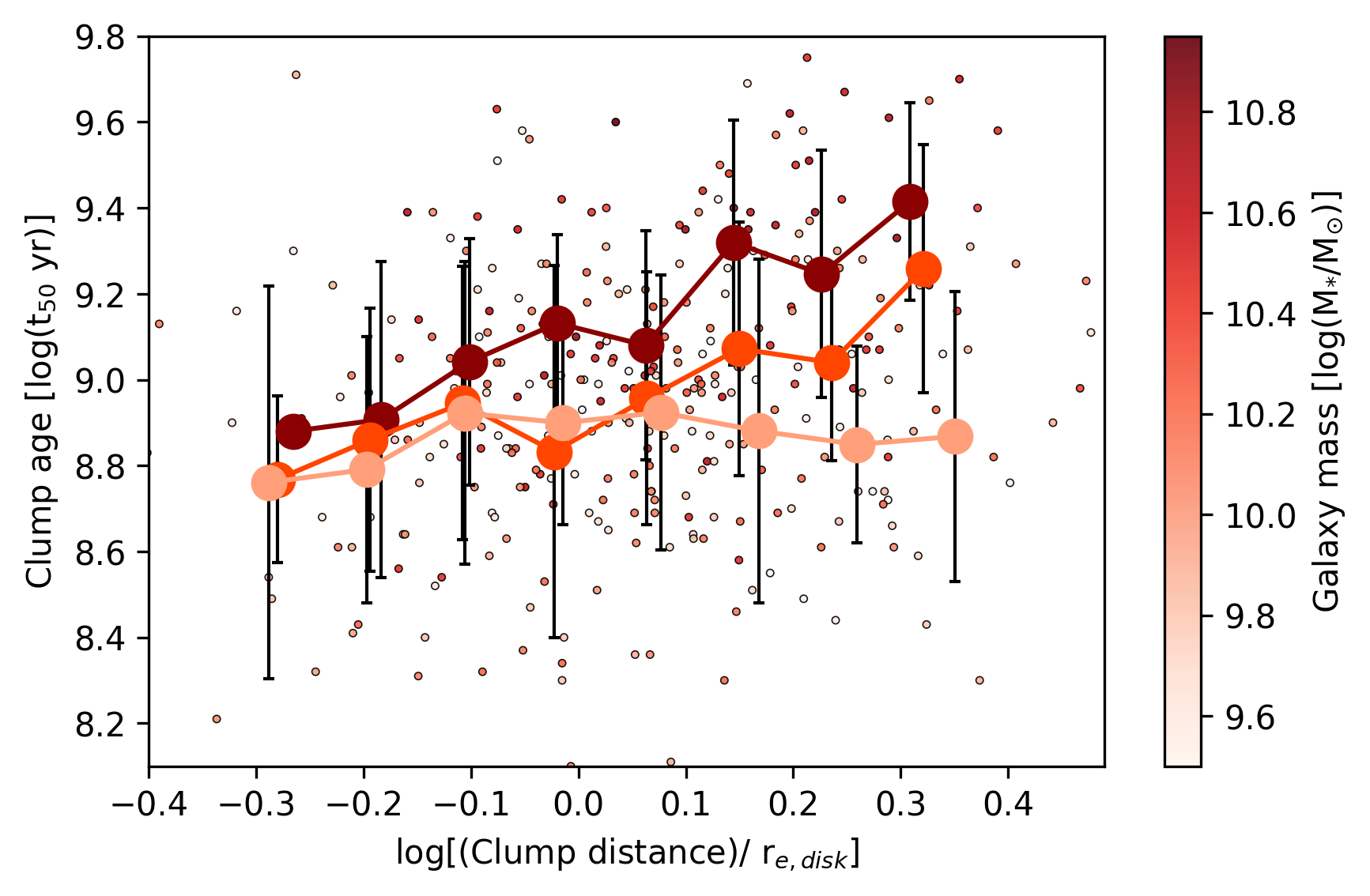}
    \includegraphics[width=0.48\textwidth]{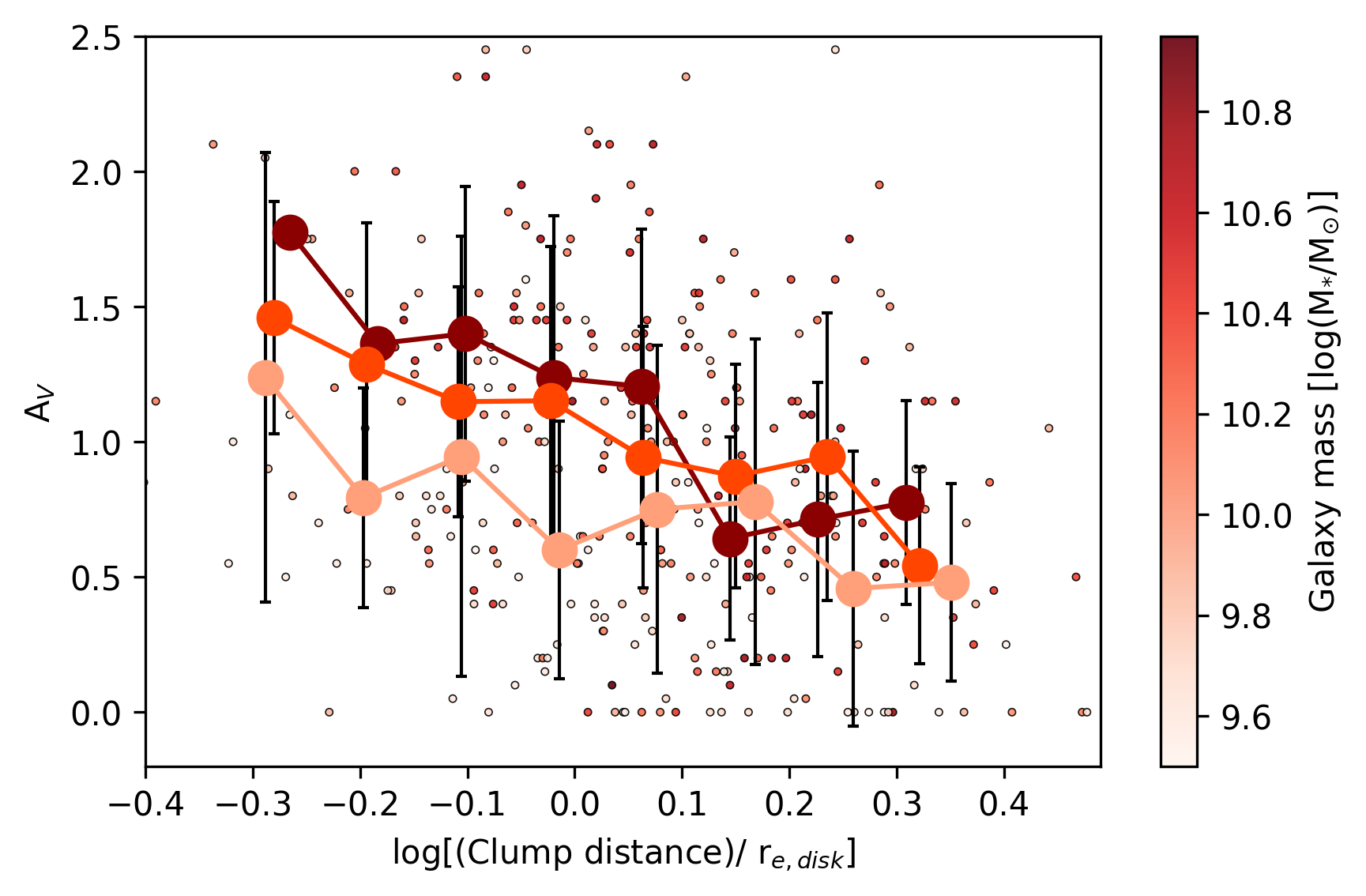}
    \caption{The SED-derived median stellar mass-weighted age (t$_{50}$) and dust attenuation (A$_{V}$), on the left and right respectively, as a function of distance from the galactic core across our sample.}
    \label{fig:sed_age_av}
\end{figure*}

As discussed in the main text, the measured values of the SFR, stellar age and dust attenuation (A$_{V}$) would not be accurate enough to draw reliable conclusions simply given the access to only wide-band photometric data up to rest frame near-IR. Nevertheless, we provide the variation of these properties as a function of distance from the centre of galaxies across our sample (Appendix Fig.~\ref{fig:sed_sfr}, \ref{fig:sed_age_av}). 

\bsp	
\label{lastpage}
\end{document}